\documentclass[prd,twocolumn,nofootinbib,showpacs,amsmath,amssymb,floatfix,eqsecnum]{revtex4-1}

\usepackage{amsmath}
\usepackage{amssymb}
\usepackage{amsthm}
\usepackage{amsfonts}

\usepackage{graphicx,color,framed}
\usepackage{hyperref}
\usepackage{times}
\usepackage{enumerate}
\usepackage{lipsum}
\usepackage{slashed}
\usepackage{url}
\usepackage{comment}
\usepackage{bbm}

\hypersetup{
    colorlinks=true, 
    linktoc=all,     
    linkcolor=blue,  
}

\newcommand{\<}{\langle}
\renewcommand{\>}{\rangle}
\newcommand{\be}{\begin{equation} }
\newcommand{\ee}{\end{equation} }
\newcommand{\ba}{\begin{eqnarray} }
\newcommand{\ea}{\end{eqnarray} }

\newcommand{\bpm}{\begin{pmatrix}}
\newcommand{\epm}{\end{pmatrix}}
\newcommand{\bmm}{\begin{matrix}}
\newcommand{\emm}{\end{matrix}}

\newcommand{\bea}{\begin{eqnarray}}
\newcommand{\eea}{\end{eqnarray}}

\newcommand{\beq}{\begin{equation} }
\newcommand{\eeq}{\end{equation} }
\newcommand{\beqs}{\begin{equation} \begin{split}}
\newcommand{\eeqs}{\end{split} \end{equation} }
\newcommand{\es}{\end{split}}

\begin{document}

\title{Anomalies in bosonic SPT edge theories: connection to $F$-symbols and a method for calculation}

\author{Kyle Kawagoe}
\author{Michael Levin}
\affiliation{Department of Physics, Kadanoff Center for Theoretical Physics, University of Chicago, Chicago, Illinois 60637,  USA}

\begin{abstract}
We describe a systematic procedure for determining the identity of a 2D bosonic symmetry protected topological (SPT) phase from the properties of its edge excitations. Our approach applies to general bosonic SPT phases with either unitary or antiunitary symmetries, and with either continuous or discrete symmetry groups, with the only restriction being that the symmetries must be on-site. Concretely, our procedure takes a bosonic SPT edge theory as input, and produces an element $\omega$ of the cohomology group $H^3(G, U_T(1))$. This element $\omega \in H^3(G, U_T(1))$ can be interpreted as either a label for the bulk 2D SPT phase or a label for the anomaly carried by the SPT edge theory. The basic idea behind our approach is to compute the $F$-symbol associated with domain walls in a symmetry broken edge theory; this domain wall $F$-symbol is precisely the anomaly we wish to compute. We demonstrate our approach with several SPT edge theories including both lattice models and continuum field theories.
\end{abstract}

\maketitle


\section{Introduction}
\label{Intro}

A gapped quantum many-body system belongs to a nontrivial ``symmetry-protected topological'' (SPT) phase if it has two properties: (i) the ground state is unique and short-range entangled\footnote{A state $|\Psi\>$ is ``short-range entangled'' if it can be transformed into a product state by local unitary transformation, i.e. a unitary of the form $U = \mathcal{T} \exp( -i\int_0^T H(t) dt)$ where $H$ is a local Hermitian operator.}; and (ii) it is not possible to adiabatically connect the system to another system with a trivial (product-state) ground state without breaking one or more global symmetries~\cite{PollmannSPT,FidkowskiKitaevSPT,ChenGuWenSPT,SchuchSPT,ChenGuWenSPTcomplete,SPTCo}. Famous examples of SPT phases include the 2D and 3D topological insulators~\cite{HasanKaneRMP,QiZhangRMP} and the 1D Haldane spin-$1$ chain~\cite{HaldaneChain}.

The most important physical property of nontrivial SPT phases is that these systems support robust gapless boundary modes. These boundary modes are ``protected'' in the sense that they cannot be gapped out without breaking one or more global symmetries~\cite{KaneMele,XuMoore,WuBernevigZhang,ChenMatrix,LevinGu,Else} or, in the case of 3D or higher dimensional systems, introducing topological order on the boundary~\cite{VishwanathSenthil,MetlitskiKaneFisherTI,ChenFidkowskiVishwanath,WangPotterSenthil,bonderson13}.  

A basic question is how to determine the identity of an SPT phase from the properties of its boundary modes. This ``bulk-boundary correspondence'' is largely understood for non-interacting fermionic SPTs. For example, in the case of 2D time-reversal symmetric insulators, one can determine whether the bulk phase is a trivial insulator or a topological insulator based on whether there are an even or odd number of Kramers pairs of edge modes. Similar formulas expressing bulk invariants in terms of boundary modes are known for other non-interacting fermion SPTs~\cite{HasanKaneRMP,QiZhangRMP}. The interacting case, however, is less understood, especially in two and higher dimensions. This paper seeks to address the interacting problem in the case of 2D bosonic SPT phases with on-site (i.e. non-spatial) symmetries. We ask: how can one determine the identity of a 2D bosonic SPT phase from its (1D) edge modes?

To make this question more concrete, let us recall the conjectured cohomology classification of 2D bosonic SPT phases~\cite{SPTCo}. According to this classification, there is a one-to-one correspondence between 2D bosonic SPT phases with on-site symmetry group $G$ and elements of the cohomology group $H^3(G,U_T(1))$. Our problem is thus to compute an element $\omega \in H^3(G, U_T(1))$ from a bosonic SPT edge theory. This element $\omega \in H^3(G, U_T(1))$ can be interpreted as describing the \emph{anomaly} carried by the edge theory.

Significant progress on this problem has been made in previous work. In a pioneering paper, Chen, Liu, and Wen~\cite{ChenMatrix} showed how to compute the anomaly $\omega \in H^3(G,U(1))$ for any bosonic SPT edge theory whose symmetries are represented by matrix product unitary operators. Later, in another important advance, Else and Nayak~\cite{Else} introduced a method for computing anomalies based on the idea of spatially restricting symmetry operators. This symmetry restriction method applies to any SPT edge theory whose symmetries are local unitary transformations, i.e. of the form $U = \mathcal{T} \exp( -i\int_0^T H(t) dt)$ for some local Hermitian operator $H$. In another line of research, several authors have presented approaches for determining anomalies in SPT edge theories with conformal symmetry, using orbifold~\cite{SuleRyuorbifold, Bultinckanomaly, TiwariRyubbc, linshao} or orientifold constructions~\cite{HsiehRyuorientifold}, though it is not obvious how to extend the latter approaches to general symmetry groups.

One limitation of the approaches introduced in Ref.~\onlinecite{ChenMatrix}, \onlinecite{Else} is that they do not apply to SPT edge theories with antiunitary symmetries except in special cases~\cite{Else}. Another limitation is that they require that the symmetries take a particular form -- either a matrix product operator or a local unitary transformation. A priori there could be edge theories where the symmetries cannot be written in these forms, or it may be difficult to write explicitly.

In this paper, we present an alternative approach for computing anomalies in bosonic SPT edge theories, which addresses these issues. Unlike previous work, our approach applies to general bosonic SPT edge theories with both unitary and antiunitary symmetries. Our only restriction is that the underlying 2D symmetry must be on-site.

The basic idea behind our approach is simple. First, we choose an edge Hamiltonian that spontaneously\footnote{Alternatively, we can break the symmetry explicitly rather than spontaneously. See Sec.~\ref{sec:continuous} for details.} breaks all the symmetries on the edge and opens up an energy gap. Such a Hamiltonian has a collection of degenerate ordered ground states related to one another by symmetry transformations. The elementary excitations are domain walls between the different ground states. These domain walls can be fused together to form new domain walls, much like anyon excitations in 2D topological systems. This allows one to define an ``$F$-symbol'' that describes the phase difference associated with fusing domain walls in different orders. This domain wall $F$-symbol is precisely the anomaly we wish to compute: we show that $F$ is naturally an element of $H^3(G, U_T(1))$, and that $F$ depends only on the edge theory and not on other details. Some care is required to compute $F$, but we describe a concrete procedure for performing this computation using the formalism of Ref.~\onlinecite{KL}. We note that the connection between domain wall $F$-symbols and anomalies was also alluded to in Ref.~\onlinecite{RooseBultinckdw} in the context of a $\mathbb{Z}_2$ SPT edge theory.

This paper is organized as follows. In Sec.~\ref{sec:setup} we explain the basic setup for our problem. Then, in Sec.~\ref{sec:unitary}, we present our anomaly computation procedure in the simplest case: SPT edge theories with discrete unitary symmetry groups. We illustrate our procedure with several (discrete unitary) examples in Sec.~\ref{sec:examples}. In Sec.~\ref{sec:conn_symm_rest}, we discuss the connection between our procedure and the symmetry restriction approach of Ref.~\onlinecite{Else}, and we prove that the two approaches agree with one another in cases where both are applicable. In Sec.~\ref{sec:antiunitary} we consider the general case of SPT edge theories with both unitary and antiunitary symmetries, and we show that the same procedure works in this case. Sec.~\ref{sec:continuous} discusses how to extend our approach to the case of \emph{continuous} symmetries. Finally, we give our conclusions in Sec. \ref{sec:conclusion}. Technical details are discussed in the Appendix.

\section{Setup}
\label{sec:setup}
We begin by defining what we mean by an ``edge theory'', or more precisely, a bosonic SPT edge theory. At an intuitive level, a bosonic SPT edge theory is a collection of data that describes the low energy edge excitations of an SPT phase. More specifically, a bosonic SPT edge theory consists of three pieces of data: 
\begin{enumerate}
\item{A Hilbert space $\mathcal{H}$.}
\item{A complete list of ``local operators'' $\{\mathcal{O}\}$ acting in $\mathcal{H}$.}
\item{A collection of (unitary or anti-unitary) symmetry transformations $\{U^g : g \in G\}$ acting on $\mathcal{H}$.}
\end{enumerate}
Each of these pieces of data has a simple physical interpretation: the Hilbert space $\mathcal{H}$ describes the subspace of low energy edge excitations; the list $\{\mathcal{O}\}$ describes the low energy projection of local operators in the original 2D system; and the $\{U^g\}$ operators describe how the edge excitations transform under the symmetry.

In order to qualify as a valid bosonic SPT edge theory, we require that the above data is physically \emph{realizable} as the edge of some 2D SPT Hamiltonian with on-site symmetries. That is, we require the existence of a 2D Hamiltonian $H_{2D}$ that belongs to an SPT phase with (on-site) symmetry group $G$ and that has the following properties:
\begin{enumerate}
\item{The Hilbert space $\mathcal{H}$ is isomorphic to the subspace of low energy edge excitations of $H_{2D}$.}

\item{The operators $\{\mathcal{O}\}$ correspond to local operators of the 2D system, projected into this low energy subspace.}

\item{The $\{U^g\}$ transformations describe how the low energy subspace transforms under the symmetries in $G$.} 
\end{enumerate}
See Secs.~\ref{sec:examples}, \ref{sec:z2z2example}, \ref{sec:IQH_example} for examples of bosonic SPT edge theories. In general, bosonic SPT edge theories can be described using either continuum fields or lattice degrees of freedom and our results apply equally well to both cases. 

\begin{figure}[t]
\centering
\includegraphics[width=1.0\columnwidth]{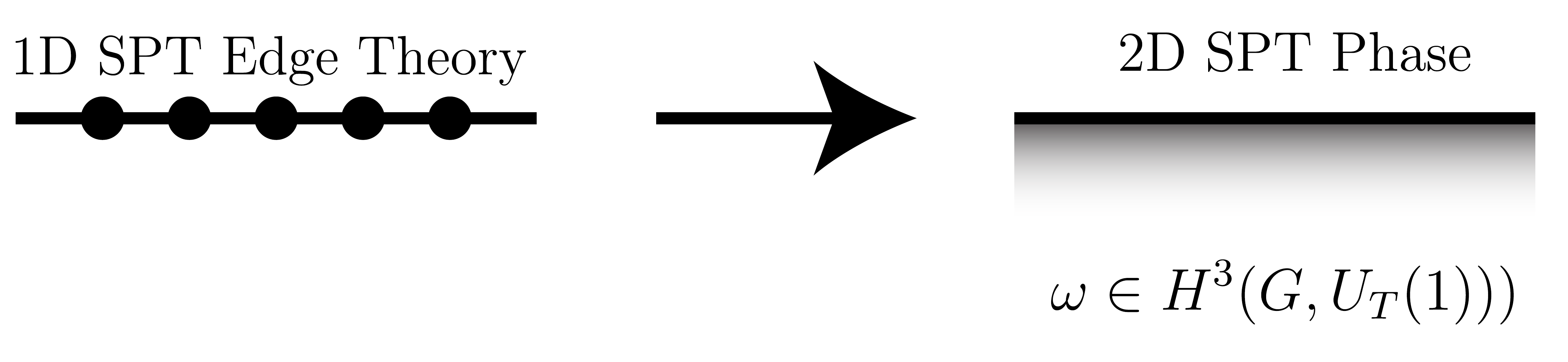}
\caption{Schematic picture for our main result: we describe a procedure that takes a 1D bosonic SPT edge theory as input, and produces as output an element $\omega \in H^3(G,U_T(1))$. This element $\omega$ can be interpreted as either a label for the bulk 2D SPT phase or the anomaly carried by the edge theory.}
\label{fig:flow_chart}
\end{figure}

With this background, we can now state our main result, (summarized in Fig.~\ref{fig:flow_chart}): we describe a systematic procedure that takes a bosonic SPT edge theory as input, and that outputs an element $\omega$ of the cohomology group $H^3(G, U_T(1))$  (see Sec.~\ref{sec:antiunitary_cohom} for a definition). As explained in the introduction, the element $\omega \in H^3(G, U_T(1))$ can be interpreted in two equivalent ways: $\omega$ can be thought of as either a label for the bulk 2D SPT phase or a label for the \emph{anomaly} carried by the SPT edge theory. Thus, depending on one's point of view, our procedure provides a systematic method for computing either the bulk 2D SPT phase or the anomaly associated with a given edge theory.

\section{Discrete unitary symmetries}
\label{sec:unitary}

\subsection{Outline of procedure}
\label{sec:unitary_outline}

In this section, we outline our procedure in the case where the symmetry group $G$ is discrete and unitary.

We start by reviewing the definition of the cohomology group $H^3(G, U(1))$, since this is the group that describes the output of our procedure in the case of discrete unitary symmetries. The cohomology group $H^3(G, U(1))$ consists of equivalence classes of functions $\omega:G\times G\times G\rightarrow U(1)$ obeying the condition
\begin{align}
\label{cocycle1}
\frac{\omega(g,h,k) \omega(g,hk,l) \omega(h,k,l)}{\omega(gh,k,l) \omega(g,h,kl)} = 1.
\end{align}
Any function $\omega$ obeying (\ref{cocycle1}) is called a ``cocycle.'' The equivalence relation between different coycles is defined as follows: $\omega \equiv \omega'$ if $\omega'/\omega$ is a ``coboundary,'' i.e.
\begin{align}
\label{coboundary}
\frac{\omega'(g,h,k)}{\omega(g,h,k)} = \frac{\nu(gh,k)\nu(g,h)}{\nu(g,hk)\nu(h,k)}
\end{align}
for some function $\nu:G\times G\rightarrow U(1)$. We will see that the output of our procedure is a cocycle $\omega$, which is uniquely defined up to multiplication by a coboundary (\ref{coboundary}); in this way, our procedure produces an element of $H^3(G,U(1))$. 

With this background, we now move on to explain our procedure. The first step in our procedure is to choose an (edge) Hamiltonian $H$ that acts within the Hilbert space $\mathcal{H}$. The Hamiltonian $H$ can be arbitrary as long as it has three properties: (i) $H$ is local, i.e. $H$ is built out of the local operators $\{\mathcal{O}\}$ from Sec.~\ref{sec:setup}; (ii) $H$ has an energy gap; and (iii) $H$ breaks the $G$-symmetry spontaneously and completely.\footnote{Actually, such a Hamiltonian is not strictly necessary for our procedure: all that we really need is a single (edge) state $|\Psi\>$ that \emph{explicitly} breaks all symmetries. See Sec.~\ref{sec:continuous} for more details.}

An important aspect of the Hamiltonian $H$ is that it has multiple degenerate ground states due to the spontaneously broken symmetry. More specifically, $H$ has $|G|$ degenerate, short-range correlated ground states, which are permuted amongst themselves by the symmetry transformations. These ground states can be naturally labeled by group elements, $g \in G$. To do this, we pick one of the degenerate ground states and denote it by $|\Omega;1\>$. We then label the other states by $\{|\Omega; g\>\}$ where $|\Omega; g\>$ is defined by
\begin{align}
|\Omega;g\>=U^g|\Omega;1\>
\label{omegagdef}
\end{align}
By construction, $U^g |\Omega; h\> = |\Omega; gh\>$.

\begin{figure}[t]
\centering
\includegraphics[width=0.8\columnwidth]{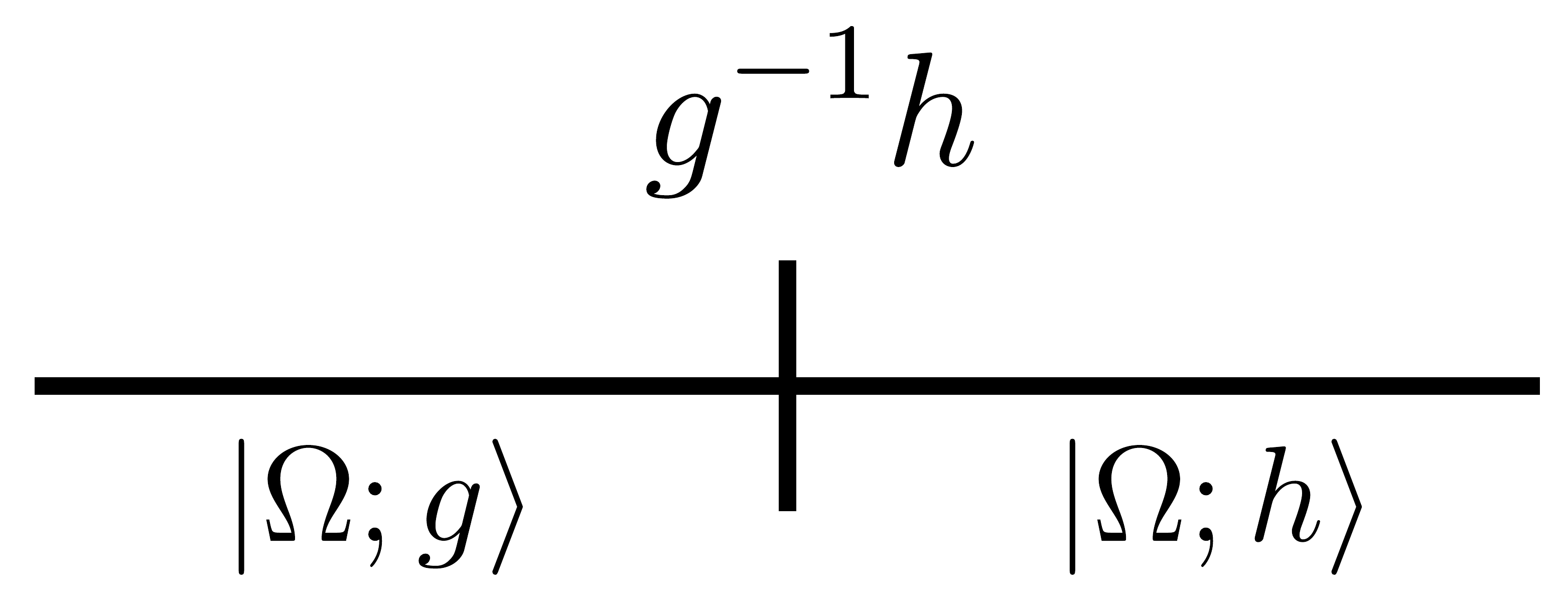}
\caption{A domain wall excitation: a state that shares the same local expectation values as one ground state $|\Omega ;g \>$ to the left of some point $x$, and another ground state $|\Omega ;h\>$ to the right of $x$. We label such a domain wall by the group element $g^{-1} h$.}
\label{fig:single_wall}
\end{figure}

Having found the ground states of $H$, the next step in our procedure is to construct domain wall excitations. We give a precise definition of domain wall excitations in Sec.~\ref{sec:micro}, but roughly speaking a domain wall excitation is a state in $\mathcal{H}$ that ``looks like'' one ground state $|\Omega; g\>$ to the left of some point $x$, and like another ground state $|\Omega; h\>$ to the right of $x$, and that interpolates between the two ground states in some arbitrary way in the vicinity of $x$ (Fig.~\ref{fig:single_wall}). Like the ground states, these domain wall excitations can be naturally labeled by group elements: in particular, we will label a domain wall with the above structure with the group element $g^{-1} h$. An important property of this labeling is that it is invariant under any global symmetry transformation $U^k$: under such a transformation, $|\Omega; g\> \rightarrow |\Omega; kg\>$ and $|\Omega; h\> \rightarrow |\Omega; kh\>$ so $g^{-1} h \rightarrow (kg)^{-1} (kh) = g^{-1} h$.

\begin{figure}[t]
\centering
\includegraphics[width=0.8\columnwidth]{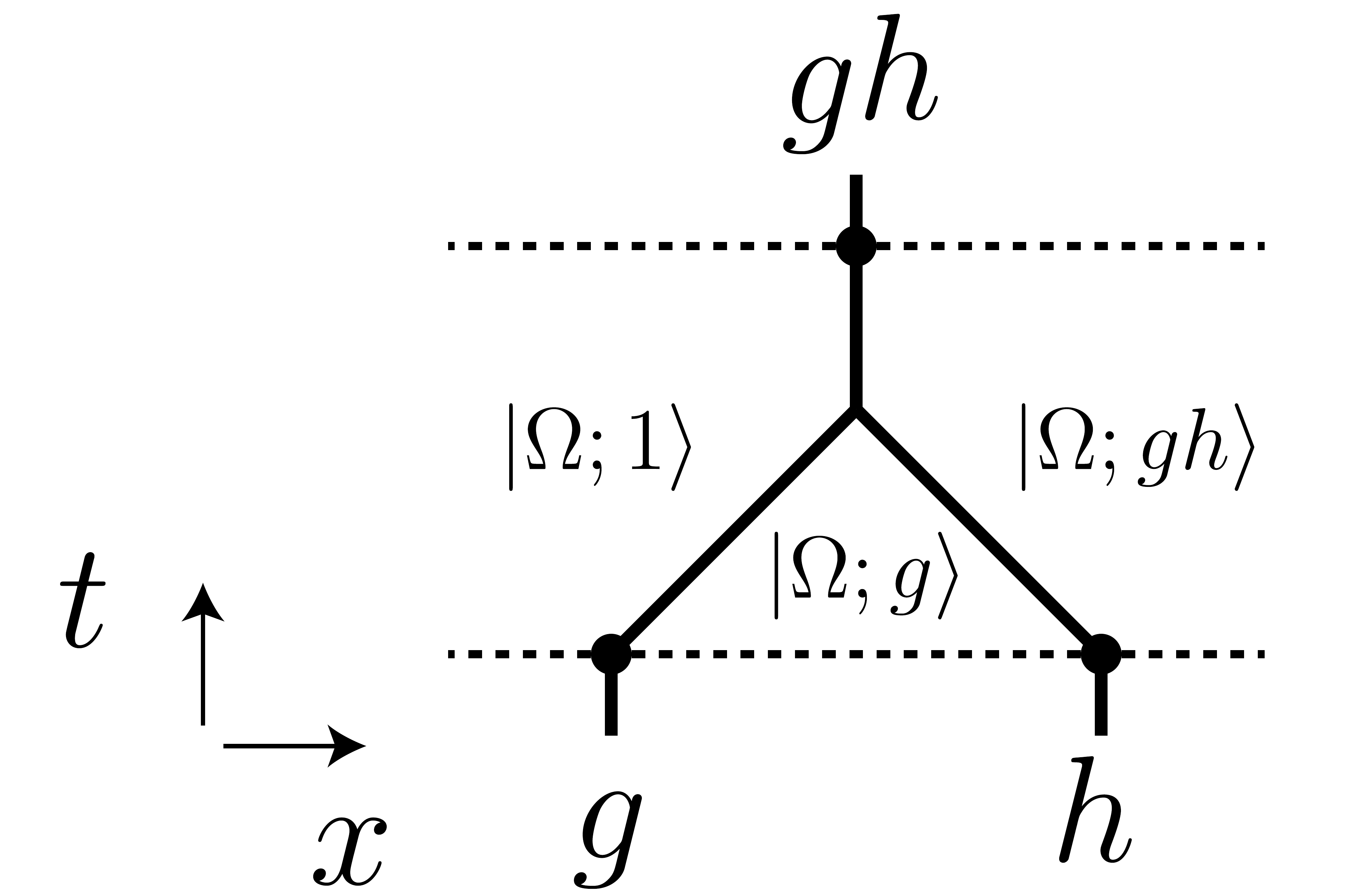}
\caption{Spacetime diagram of fusion of a $g$ and $h$ domain wall into a $gh$ domain wall. The initial state consists of three different ground states $|\Omega;1\>,|\Omega;g\>,|\Omega;gh\>$ in three different regions, separated by domain walls $g,h$. During the fusion process, the $|\Omega;g\>$ region disappears and we are left only with the $|\Omega;1\>,|\Omega;gh\>$ regions separated by a $gh$ domain wall. Reversing the arrow of time corresponds to a ``splitting" operation.}
\label{fig:fusion}
\end{figure}

Like other topological defects, domain walls cannot be created or annihilated individually. Instead, the most basic process that one can perform on domain walls is to combine them together in a process called ``fusion'': if one has a $g$ type domain wall located nearby and to the left of an $h$ type domain wall, then one can convert this pair of domain walls into a single $gh$ type domain wall by applying a local operator acting on both domain walls. This process is shown in Fig.~\ref{fig:fusion}. Alternatively, one can perform the reverse (``splitting'') process and apply a local operator that turns a $gh$ type domain wall into a $g$ type domain wall to the left of an $h$ type domain wall. 

\begin{figure}[tb]
\centering
\includegraphics[width=.99\columnwidth]{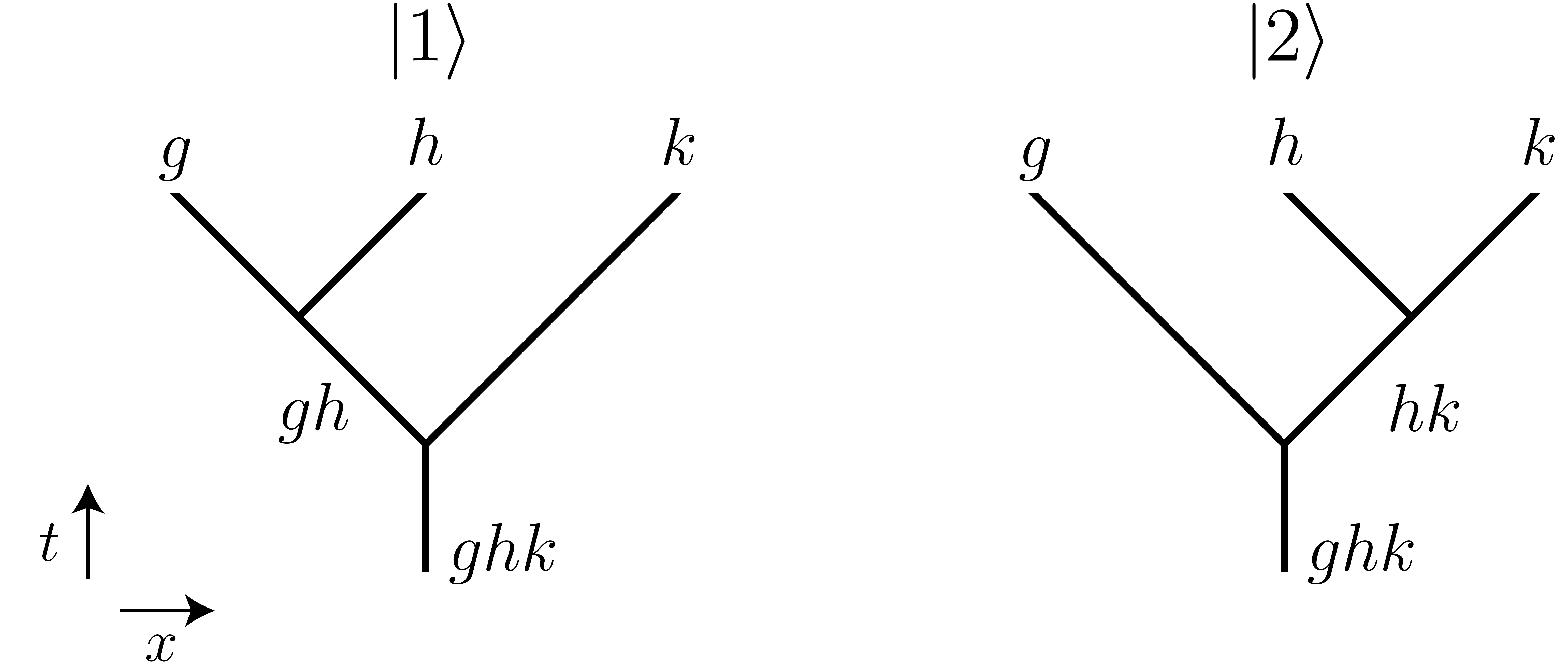}
\caption{Spacetime diagrams for two processes in which a domain wall $ghk$ splits into domain walls $g,h,k$. The final states, $|1\>, |2\>$, are equal up to the $U(1)$ phase, $F(g,h,k)$.}
\label{fig:sketch}
\end{figure}

This ability to fuse or split domain walls allows us to define an ``$F$-symbol'' for these excitations. The $F$-symbol is usually discussed in the context of 2D anyon theories~\cite{kitaevlongpaper,preskilltqc}, but it is a more general concept that can be defined for any point-like mobile defects. In particular, we can sensibly define an $F$-symbol for domain walls in 1D edge theories. The basic idea is to consider two different physical processes in which a domain wall of type $ghk$ splits into three domain walls of types $g, h, k$ (Fig. \ref{fig:sketch}). In one process, $ghk$ splits into $gh$ and $k$, and then $gh$ splits into $g$ and $h$; in the other process, $ghk$ splits into $g$ and $hk$ and then $hk$ splits into $h$ and $k$. By construction, the final states $|1\>, |2\>$ produced by these processes contain the same domain walls, $g, h, k$, at the same three positions. Therefore, the two final states $|1\>, |2\>$ must be the same up to a phase. The $F$-symbol, $F(g,h,k)$, is the phase difference between the two states: 
\begin{align}
|1\> = F(g,h,k) |2\>.
\label{Fdefab}
\end{align}

The final step in our procedure is to compute this $F$-symbol for domain wall excitations. This $F$-symbol defines our cocycle $\omega \in H^3(G, U(1))$:
\begin{align}
\omega(g,h,k) \equiv F(g,h,k)
\label{omegaF}
\end{align}

\begin{figure}[tb]
\centering
\includegraphics[width=1.0\columnwidth]{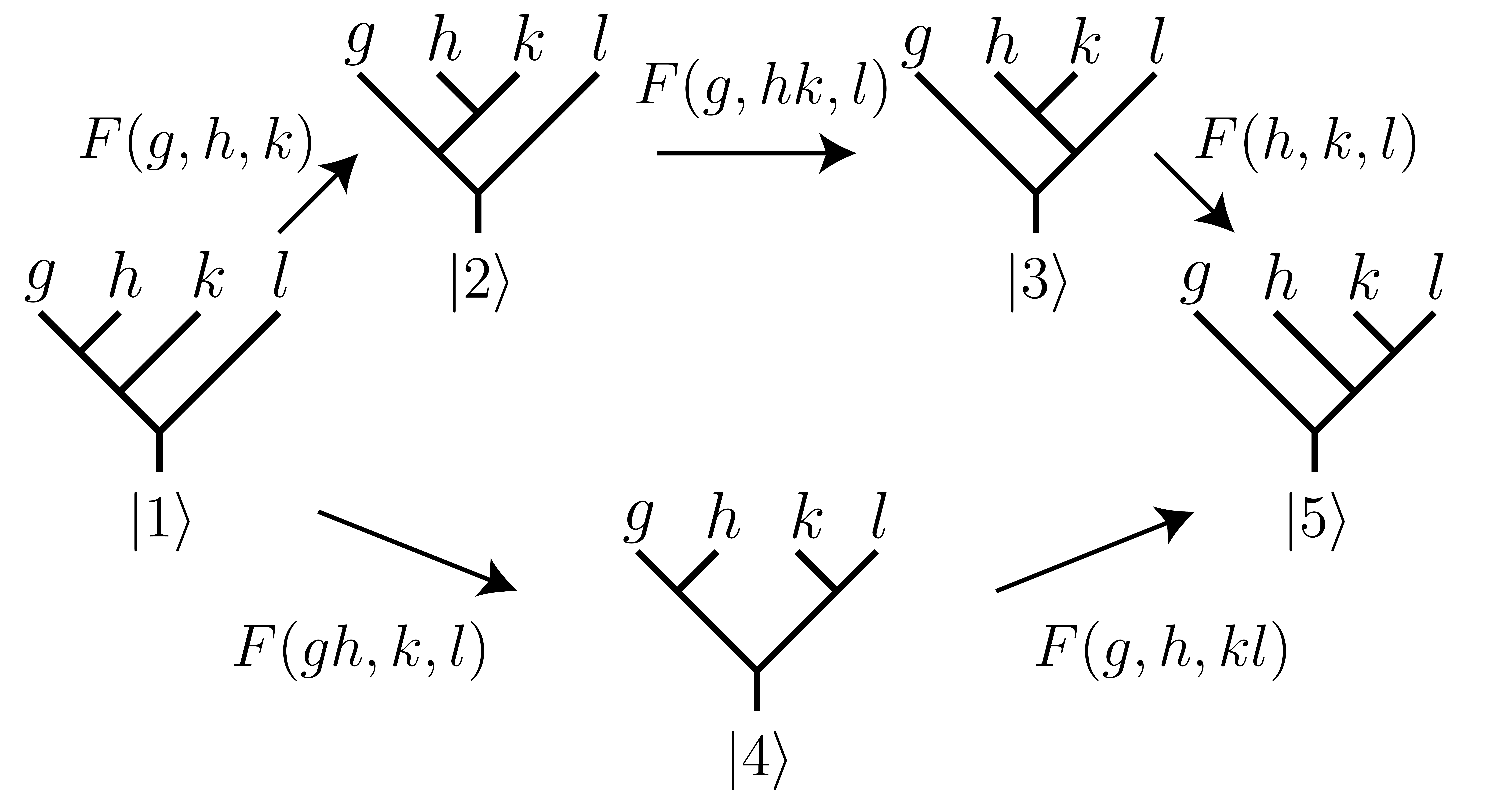}
\caption{The pentagon identity: consistency requires that the product of the three $F$-symbols on the upper path is equal to the product of the two $F$-symbols on the lower path.}
\label{fig:pentagon0}
\end{figure}

To justify Eq.~\ref{omegaF}, we need to explain why $F(g,h,k)$ is an element of $H^3(G, U(1))$. This follows from two important properties of $F$ which we prove in Sec.~\ref{sec:check_micro}. The first property is that $F$ satisfies a non-trivial constraint, known as the ``pentagon identity'':
\begin{equation}
F(g,h,k) F(g,hk,l) F(h,k,l) = F(gh,k,l) F(g,h,kl)
\label{pentid}
\end{equation}
To understand the origin of this identity, consider the $5$ processes shown in Fig.~\ref{fig:pentagon0}. Notice that the final states produced by these processes, namely $\{|1\>,...,|5\>\}$, are all the same up to a phase. We can compute the phase difference between states $|1\>$ and $|5\>$ in two different ways. In the first way, we compute the relative phases between $(|1\>, |2\>)$, $(|2\>, |3\>)$, and $(|3\>, |5\>)$ using (\ref{Fdefab}); in the second way, we compute the relative phases between $(|1\>, |4\>)$ and $(|4\>, |5\>)$. Demanding consistency between the two calculations gives the pentagon identity (\ref{pentid}). 

The second property of $F$ is that is has an inherent ambiguity: it is only well-defined up to transformations of the form
\begin{equation}
F(g,h,k)\rightarrow F(g,h,k)\frac{\nu(gh,k)\nu(g,h)}{\nu(g,hk) \nu(h,k)}
\label{gaugetrans}
\end{equation} 
where $\nu(g,h) \in U(1)$. To understand where this ambiguity comes from, it is helpful to think about the physical processes in Fig.~\ref{fig:sketch} as being implemented by a sequence of two ``splitting operators'' applied to an initial state $|ghk_1\>$. The key point is that the phases of these splitting operators are arbitrary. If we multiply the four splitting operators by four phases, $\nu(gh,k), \nu(g,h), \nu(g,hk), \nu(h,k)$, this changes $F$ by exactly the above transformation (\ref{gaugetrans}). We will call the transformations in (\ref{gaugetrans}) ``gauge transformations.''

Comparing the above properties of $F$ to the definition of $H^3(G,U(1))$, we can see that the pentagon identity (\ref{pentid}) is identical to the cocycle condition (\ref{cocycle1}), while the ambiguity (\ref{gaugetrans}) is equivalent to the equivalence relation (\ref{coboundary}) on cocycles. Hence, $F$ naturally an element of $H^3(G,U(1))$, as we claimed earlier.

\subsection{Microscopic definition of $F$-symbol for domain walls}
\label{sec:micro}

In this section, we give precise, operational definitions of domain wall states and their associated $F$-symbols. These definitions are essential for making our procedure a useful calculational tool; they are also important for putting our procedure on a firm foundation. We note that these definitions closely parallel the microscopic definition of the anyonic $F$-symbol that was given in Ref.~\onlinecite{KL} in the context of anyon theories.
 
To begin, let $\ell$ be a distance that is much greater than the correlation length $\xi$ of the system. The length scale $\ell$ will play an important role in the following discussion. In particular, we will only consider states in which domain wall excitations are separated by distances of at least $\ell$, and we will neglect finite size effects of order $e^{-\ell/\xi}$.

We define domain wall states as follows. For each point $x$ on the edge, we choose a state that has the same expectation values as $|\Omega;1\>$ for local operators $\{\mathcal{O}\}$ supported to the left of $x-\ell$ and the same expectation values as $|\Omega;g\>$ for local operators $\{\mathcal{O}\}$ to the right of $x+\ell$. We denote this state by $|g_x\>$. In our language, the state $|g_x\>$ describes a single domain wall of ``type $g$'' at position $x$. Note that the definition of $|g_x\>$ involves an arbitrary choice of a state: we will show that our results do not depend on this choice. 

Next, for each domain wall state $|g_x\>$, we define a collection of symmetry partner states $|g_x; h\>$, with $h \in G$, by
\begin{align}
|g_x;h\>=U^h|g_x\>
\end{align}
By construction, $|g_x;h\>$ has the same expectation values as $|\Omega;h\>$ for local operators supported to the left of $x-\ell$ and the same expectation values as $|\Omega;hg\>$ for local operators to the right of $x+\ell$.

\begin{figure}[tb]
\centering
\includegraphics[width=1.0\columnwidth]{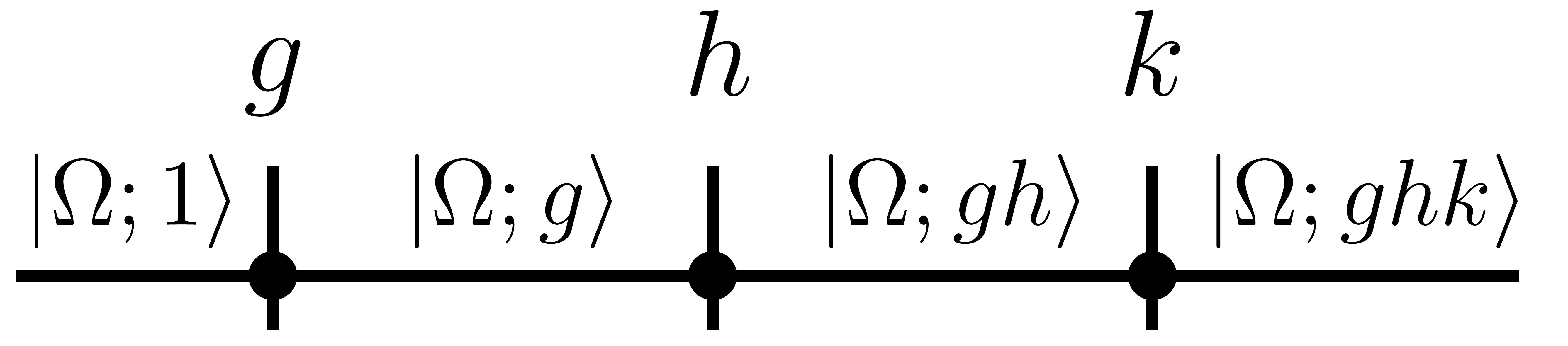}
\caption{A multi-domain wall state $|g_x, h_{x'}, k_{x''}\>$ consisting of a $g$-type, $h$-type, and $k$-type domain wall at positions $x, x', x''$. The domain walls separate four regions which share the same local expectation values as the four ground states $|\Omega;1\>,|\Omega;g\>,|\Omega;gh\>,|\Omega;ghk\>$.}
\label{fig:multidomain}
\end{figure}

We now introduce \emph{multi}-domain wall states.  Let $x_1<x_2<...<x_n$ be well-separated (i.e. having a spacing of at least $\ell$) and pick group elements $g^{(1)},g^{(2)},...,g^{(n)}\in G$. We will use the notation $|g^{(1)}_{x_{1}},g^{(2)}_{x_{2}},...\>$ to denote the multi-domain wall state that has a domain wall of type $g^{(i)}$ at each location $x_i$ and that has the same expectation values as $|\Omega;1\>$ for local operators supported to the left of all the domain walls. More precisely, we 
define $|g^{(1)}_{x_{1}},g^{(2)}_{x_{2}},...\>$ to be the unique state with the following two properties: first, for any local operator $\mathcal{O}$ supported near one domain wall $x_i$,
\begin{align}
\label{domain_wall_inv}
\<g^{(1)}_{x_{1}},g^{(2)}_{x_{2}},...|\mathcal{O}|g^{(1)}_{x_{1}},g^{(2)}_{x_{2}},...\>=\<g_{x_i}^{(i)};g_L|\mathcal{O}|g_{x_i}^{(i)};g_L\>
\end{align}
where $g_L$ is the product of all domain wall types to the left of $\mathcal{O}$:
\begin{align}
\label{gL}
g_L=g^{(1)}\cdots g^{(i-1)}
\end{align}
Second, for any operator $\mathcal{O}$ supported away from the domain walls,
\begin{align}
\<g^{(1)}_{x_{1}},g^{(2)}_{x_{2}},...|\mathcal{O}|g^{(1)}_{x_{1}},g^{(2)}_{x_{2}},...\>=\<\Omega;g_L|\mathcal{O}|\Omega;g_L\>
\end{align}
where $g_L$ is again the product of domain wall types to the left of $\mathcal{O}$.  
These two properties imply that multidomain wall states have a structure like that in Fig.~\ref{fig:multidomain}.

An important corollary of Eq.~\ref{domain_wall_inv}, which we will need below, is that for any operator $\mathcal{O}$ that is supported near a domain wall $g$ at point $x$ and is invariant under all the symmetries (i.e. $U^h \mathcal{O} (U^h)^{-1} = \mathcal{O}$), the following identity holds:
\begin{align}
\label{sym_inv}
\<...,g_x,...|\mathcal{O}|...,g_x,...\>=\<g_x|\mathcal{O}|g_x\>
\end{align}

Now that we have fixed our definitions of domain wall states, the next step is to define \emph{movement} operators for these domain walls. Given any group element $g \in G$, and any pair of points, $x, x'$, we say that $M^g_{x'x}$ is a movement operator if it satisfies three conditions: (i) $M^g_{x'x}$ obeys
\begin{equation}
M^g_{x'x} |g_x\> \propto |g_{x'}\>
\label{mdef1}
\end{equation}
where the proportionality constant is a $U(1)$ phase; (ii) $M^g_{x'x}$ is invariant under all the symmetries, i.e.
\begin{equation}
U^h M^g_{x'x} (U^h)^{-1} =M^g_{x'x}
\label{msymm}
\end{equation}
and (iii) $M^g_{x'x}$ is \emph{local} in the sense that it is supported in the neighborhood of the interval containing $x$, $x'$.   

Here, the symmetry condition (\ref{msymm}) is important because it guarantees that the analog of Eq.~\ref{mdef1} holds for any (single) domain wall state $|g_x; h\>$:
\begin{align}
M^g_{x'x}|g_x;h\>\propto |g_{x'};h\>
\label{aug_move}
\end{align}
Likewise, the locality condition is important because it guarantees that the analog of Eq.~\ref{mdef1} holds for any multi-domain wall state of the form $|...,g_x,...\>$:  that is,
\begin{equation}
M^g_{x'x} |...,g_x,...\> \propto |...,g_{x'},...\>
\label{mdef2}
\end{equation}
as long as the other domain walls in $|...,g_x,...\>$ are well-separated from the interval containing $x$ and $x'$.  Again the constant of proportionality is a $U(1)$ phase.

To derive Eq.~\ref{mdef2} from Eq.~\ref{mdef1}, consider the expectation value of any local operator, $\mathcal{O}$, supported in the neighborhood of $[x,x']$ (or $[x',x]$ if $x'<x$), in the two states $|...,g_{x'},...\>$ and $M^g_{x'x} |...,g_x,...\>$. Using (\ref{domain_wall_inv}) and (\ref{aug_move}), we can see that $\mathcal{O}$ has the same expectation value in the two states, $|...,g_{x'},...\>$ and $M^g_{x'x} |...,g_x,...\>$:
\begin{align}
&\<...,g_{x'},...| \mathcal{O} |...,g_{x'},...\> = \<g_{x'};g_L| \mathcal{O} |g_{x'};g_L\> \nonumber \\
& \hspace{3 cm}= \<g_{x};g_L| (M^g_{x'x})^\dagger \mathcal{O} M^g_{x'x}  |g_x;g_L\> \nonumber \\
& \hspace{3 cm} =\<...,g_{x},...| (M^g_{x'x})^\dagger \mathcal{O} M^g_{x'x}  |...,g_x,...\>
\end{align}
The two states, $|...,g_{x'},...\>$ and $M^g_{x'x} |...,g_x,...\>$ also share the same expectation values for local operators supported away from the interval $[x,x']$ (or $[x',x]$) by virtue of the short ranged correlations of these states. Therefore, by the uniqueness property of our domain wall states, we obtain Eq.~\ref{mdef2}.

In addition to the movement operators, we also define \emph{splitting} operators for our domain walls. Fix two well-separated points on the line, which we will call `$1$' and `$2$'. For any pair of domain walls $g,h$, we say that $S(g,h)$ is a splitting operator if it satisfies three conditions: (i) $S(g,h)$ satisfies
\begin{equation}
S(g,h)|gh_1\> \propto|g_1, h_2\>
\end{equation}
where the proportionality constant is a $U(1)$ phase; (ii) $S(g,h)$ is invariant under all the symmetries $U^k$; (iii) $S(g,h)$ is supported in the neighborhood of the interval $[1,2]$. 

Just as before, it can be shown that these conditions guarantee that the splitting operators can be applied to any multi-domain wall state of the form $|...,gh_1,...\>$ provided that the other domain walls are located far from the interval $[1,2]$:
\begin{equation}
S(g,h)|...,gh_1,...\> \propto|...,g_1, h_2,...\>
\label{smultiprop}
\end{equation}
where the proportionality constant is a $U(1)$ phase. Note that, unlike the movement operators, we only define splitting operators that act on a \emph{single} interval $[1,2]$ on the $x$-axis.

\begin{figure}[tb]
\centering
\includegraphics[width=0.9\columnwidth]{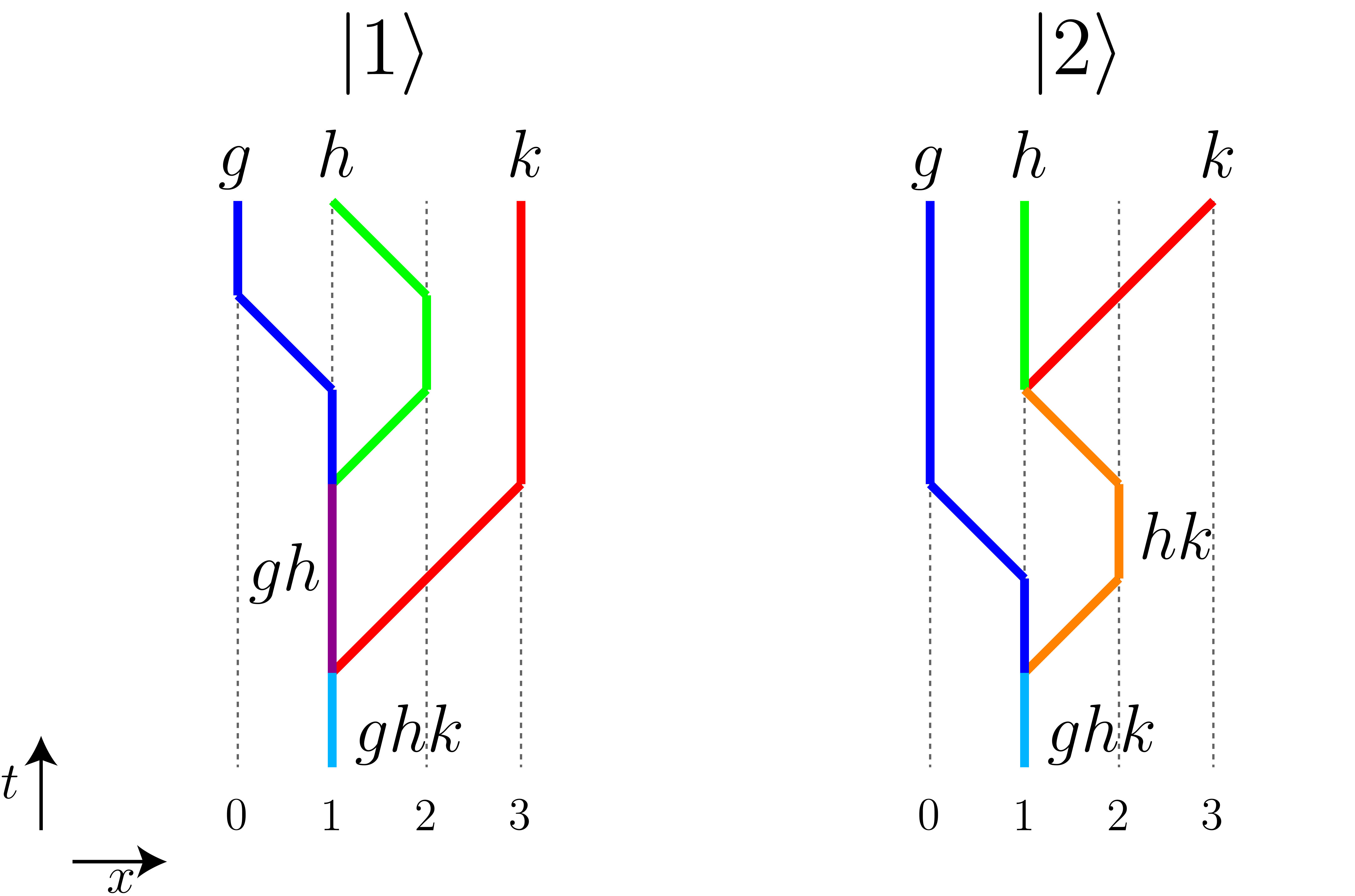}
\caption{The two processes that are compared in the microscopic definition of the domain wall $F$-symbol. Starting from an initial state $|ghk_1\>$, movement and splitting operators are applied sequentially to obtain final states $|1\>$ and $|2\>$ on the left and right, respectively. States $|1\>, |2\>$ are both proportional to $|g_0,h_1,k_3\>$. The $x$-axis is the position on the edge and the $t$-axis shows operator ordering.
}
\label{fig:two}
\end{figure}

With this setup, we are now ready to define the $F$-symbol. The first step is to fix some choice of domain wall states, $|g_x\>$, and some choice of movement and splitting operators, $M^g_{x'x}, S(g,h)$. Next, consider the initial state $|ghk_1\>$, i.e. the state with a single domain wall $ghk$ at position $1$. We then apply two different sequences of movement and splitting operators to $|ghk_1\>$, denoting the final states by $|1\>$ and $|2\>$:
\begin{align}
|1\>&= M^h_{12} M^g_{01} S(g,h) M^k_{32} S(gh,k) |ghk_1\> \nonumber \\
|2\>&= M^k_{32} S(h,k) M^{hk}_{12} M^g_{01} S(g,hk) |ghk_1\> 
\label{fdef1}
\end{align}
These two processes are shown in Fig.~\ref{fig:two}. By construction, the final states $|1\>, |2\>$ produced by these processes both contain domain walls $g, h, k$ at positions $0, 1, 3$, respectively. In particular, this means that $|1\>, |2\>$ are the same up to a phase. We define the $F$-symbol, $F(g,h,k)$, to be this phase difference:
\begin{equation}
F(g,h,k)=\<2|1\>
\label{fdef2}
\end{equation}

\subsection{Checking the microscopic definition}
\label{sec:check_micro}
To show that our microscopic definition is correct, we need to establish two properties of $F$: (i) $F$ is well-defined in the sense that different choices of domain wall states and movement and splitting operators give the same $F$ up to a gauge transformation (\ref{gaugetrans}); and (ii) $F$ obeys the pentagon equation (\ref{pentid}). We prove property (ii) in Appendix~\ref{pentagon}; the goal of this section is to prove property (i). 

As a warm-up, let us see how $F$ transforms if we only change the \emph{phase} of the movement and splitting operators. That is, suppose we replace
\begin{align}
M^g_{x'x} \rightarrow e^{i\theta_{x'x}(g)} M^g_{x'x}, \quad S(g,h) \rightarrow e^{i\phi(g,h)} S(g,h) 
\end{align}
for some real-valued $\theta, \phi$.  Substituting these transformations into (\ref{fdef1}-\ref{fdef2}) gives
\begin{align}
F(g,h,k) \rightarrow F(g,h,k) \frac{e^{i\phi(gh,k)}e^{i\phi(g,h)} e^{i\theta_{12}(h)}}{e^{i\phi(g,hk)}e^{i\phi(h,k)}e^{i\theta_{12}(hk)}}
\end{align}
Crucially, this transformation is identical to a coboundary transformation (\ref{gaugetrans}) with
\begin{equation}
\nu(g,h)=e^{i[\phi(g,h)+\theta_{12}(h)]}
\label{nuphitheta}
\end{equation}
This is exactly what we want: different phase choices lead to the same $F$, up to a coboundary, or gauge transformation. 

With this warm-up, we are now ready to consider the general case where we change the movement and splitting operators in an arbitrary way:
\begin{align}
M^g_{x'x} \rightarrow M'^g_{x'x}, \quad \quad S(g,h) \rightarrow S'(g,h)
\end{align}
To analyze this case, we note that the definition of movement and splitting operators implies that
\begin{align}
M'^g_{x'x} |g_x\> &= e^{i \theta^g_{x'x}} M^g_{x'x} |g_{x}\> \nonumber\\
S'(g,h)|gh_1\> &= e^{i \phi(g,h)} S(g,h)|gh_1\>     
\end{align}
for some real valued $\theta^g_{x'x}$ and $\phi(g,h)$. Likewise, for \emph{multi}-domain wall states,
\begin{align}
M'^g_{x'x} |...,g_x,...\> &= \omega_1 \cdot M^g_{x'x} |...,g_{x},...\> \label{maphase}\\
S'(g,h)|...,gh_1,...\> &= \omega_2 \cdot S(g,h)|...,gh_1,...\>   
\end{align}
for some $U(1)$ phases $\omega_1, \omega_2$. 

To proceed further, we need to find the relationship between the multi-domain wall phases $\omega_1, \omega_2$ and the single domain wall phases $e^{i\theta^g_{x'x}}$ and $e^{i\phi(g,h)}$. To do this, we multiply the two sides of (\ref{maphase}) by $ \<...,g_{x},...|(M^g_{x'x})^\dagger$ and then use property (\ref{sym_inv}) to derive 
\begin{align}
\omega_1 &= \<...,g_{x},...|(M^g_{x'x})^\dagger M'^g_{x'x} |...,g_x,...\> \nonumber \\
&= \<g_{x}|(M^g_{x'x})^\dagger M'^g_{x'x} |g_x\> \nonumber \\
&= e^{i \theta^g_{x'x}}
\end{align}
By the same reasoning, $\omega_2 = e^{i \phi(g,h)}$. We conclude that
\begin{align}
M'^g_{x'x} |...,g_x,...\> &= e^{i\theta_{x'x}(g)} M^g_{x'x} |...,g_{x},...\> \nonumber \\
S'(g,h)|...,gh_1,...\> &= e^{i\phi(g,h)} S(g,h)|...,gh_1,...\>
\label{msthetaphi}
\end{align}
We are now finished: substituting the above relations (\ref{msthetaphi}) into (\ref{fdef1}), we again see that $F$ changes by a gauge transformation with $\nu$ given by (\ref{nuphitheta}).

So far, we have shown that different choices of movement and splitting operators lead to the same $F$, up to a gauge transformation. Next, we need to check that different choices of edge Hamiltonians and domain wall states also lead to the same $F$, up to a gauge transformation. To investigate this issue, consider two choices of edge Hamiltonians, $H$ and $H'$, with ground states $\{|\Omega; g\>\}$ and $\{|\Omega; g\>'\}$ respectively. It seems reasonable to assume that $H$ and $H'$ can be adiabatically connected, i.e. there exists an interpolating Hamiltonian $H_s$, with $0 \leq s \leq 1$ with $H_0 = H$ and $H_1 = H'$ such that $H_s$ is local, gapped, and breaks the $G$-symmetry spontaneously and completely. Assuming the existence of such an interpolation, it then follows from the quasi-adiabatic continuation construction~\cite{HastingsWenQAC} that there exists a $G$-invariant, ``locality preserving'' unitary transformation $W$ that connects the two sets of ground states: that is,
\begin{align}
\{|\Omega; g\>'\} = \{W |\Omega; g\>\}
\label{OmegaWassum}
\end{align}
Here, when we say $W$ is ``locality preserving'', we mean that it has the following property: for any local operator $\mathcal{O}$, the operator $W \mathcal{O} W^{-1}$ is also local and is supported near $\mathcal{O}$. 

Comparing Eq.~\ref{OmegaWassum} with our labeling scheme (\ref{omegagdef}), we deduce that the two sets of ground states are related by
\begin{align}
|\Omega; g\>' = W U^k |\Omega; g\>
\label{OmegaWassum1}
\end{align}
for some $k \in G$. 

To proceed with our analysis, we will make the additional assumption that Eq.~(\ref{OmegaWassum1}) can be extended from ground states to (multi-)domain wall states. That is, we assume that, for any two choices of domain wall states, there exists a $G$-invariant, locality preserving unitary transformation $W$ such that
\begin{align}
|g^{(1)}_{x_1}, g^{(2)}_{x_2}, g^{(3)}_{x_3}, ...\>' = WU^k |g^{(1)}_{x_1}, g^{(2)}_{x_2}, g^{(3)}_{x_3}, ...\>
\label{OmegaWassum2}
\end{align}
This assumption is reasonable as long as the domain wall positions $x_1, x_2,...$ are well-separated.

With Eq.~(\ref{OmegaWassum2}), we are now in a position to compare the $F$-symbols for the two choices of domain wall states. First, we observe that we are free to choose the movement and splitting operators for the states $|g^{(1)}_{x_1}, g^{(2)}_{x_2}, g^{(3)}_{x_3}, ...\>'$ however we like, changing $F$ by, at most, a gauge transformation. The simplest choice that is consistent with (\ref{OmegaWassum2}) is
\begin{align}
M'^g_{x'x}=WM^g_{x'x}W^{-1},\quad S'(g,h)=WS(g,h)W^{-1}
\end{align}
With this choice, it is clear that $|1'\>= W U^k|1\>$, and $|2'\> = W U^k|2\>$. It follows that $F' = \<2'|1'\> = \<2|1\> = F$. Thus, we conclude that $F$ doesn't depend on the choice of the edge Hamiltonian or the choice of domain wall states. This completes the proof of property (i) above.


\section{Examples with discrete unitary symmetries}
\label{sec:examples}
In this section, we present several examples illustrating our method of calculating anomalies in bosonic SPT edge theories.

\subsection{Lattice edge theory for $\mathbb{Z}_2$ SPT phase}
\label{sec:z2lattice}

We start with one of the simplest non-trivial examples: a lattice edge theory for the bosonic SPT phase with symmetry group $G = \mathbb{Z}_2$. This edge theory was introduced in Refs.~\onlinecite{ChenMatrix,LevinGu}.

\subsubsection{Edge theory}

We begin by reviewing the structure of the edge theory. As explained in Sec~\ref{sec:setup}, an edge theory consists of three pieces of data: (1) a Hilbert space $\mathcal{H}$; (2) a set of local operators $\{\mathcal{O}\}$ that act in $\mathcal{H}$; and (3) a collection of symmetry transformations $\{U^g : g \in G\}$ acting within $\mathcal{H}$. 

The Hilbert space $\mathcal{H}$ for the $\mathbb{Z}_2$ SPT edge theory couldn't be simpler: the Hilbert space is equivalent to a one-dimensional spin-$1/2$ chain, with spins living on integer lattice sites $n \in \mathbb{Z}$. A complete orthonomal basis for this Hilbert space can be obtained by considering $\sigma^z$ eigenstates of the form
\begin{align}
\label{spin_labels}
|...,\alpha_{-1},\alpha_0,\alpha_1,...\>, \quad \quad \alpha_n\in\{+,-\}
\end{align}
where
\begin{align}
\sigma^z_n |...,\alpha_{-1}, \alpha_0, \alpha_1,...\> = \alpha_n |...,\alpha_{-1}, \alpha_0, \alpha_1,...\>
\end{align}
The local operators $\{\mathcal{O}\}$ in the $\mathbb{Z}_2$ SPT edge theory are the usual local operators in a spin-$1/2$ chain, namely products of Pauli spin operators $\{\sigma^x_n, \sigma^y_n, \sigma^z_n\}$ acting on a collection of nearby lattice sites.

We now move on to the symmetry transformations $U^g$. Denoting the symmetry group by $\mathbb{Z}_2 = \{1,s\}$, we only need to discuss the symmetry operator $U^s$, since $U^1 = \mathbbm{1}$. Denoting $U^s \equiv U$ for brevity, we can describe the symmetry operator $U$ by how it acts on the Pauli spin operators~\cite{LevinGu}: 
\begin{align}
\label{z2lattice_sym}
U \sigma^x_jU^{-1}&=-\sigma^z_{j-1}\sigma^x_j\sigma^z_{j+1} \nonumber \\
U \sigma^y_jU^{-1}&=\sigma^z_{j-1}\sigma^y_j\sigma^z_{j+1} \nonumber \\
U \sigma^z_jU^{-1}&=-\sigma^z_j 
\end{align}

We can also write down an explicit formula for $U$~\cite{LevinGu}:
\begin{equation}
U= - \prod_j i^{\frac{1-\sigma^z_j\sigma^z_{j+1}}{2}}\prod\limits_j\sigma_j^x
\label{z2lattice_sym2}
\end{equation}
Note that we will not need the above formula (\ref{z2lattice_sym2}) to compute the anomaly: our approach only requires knowing the transformation laws for local operators (\ref{z2lattice_sym}).

\subsubsection{Calculating the anomaly}
We now proceed to compute the anomaly associated with the edge theory described above. As we explained in Sec.~\ref{sec:unitary_outline}, the first step in calculating the anomaly is to choose a gapped (edge) Hamiltonian that breaks the $\mathbb{Z}_2$ symmetry spontaneously and completely. We use an Ising Hamiltonian
\begin{align}
H=-J\sum_i\sigma^z_i\sigma^z_{i+1}
\end{align}
with $J>0$. 

To see that this Hamiltonian spontaneously breaks the symmetry note that $\sigma^z$ is odd under the symmetry (\ref{z2lattice_sym2}) and has a non-zero expectation value in the two degenerate ground states $|+,+,...,+\>$ and $|-,-,...,-\>$.
Following the notation in Sec.~\ref{sec:unitary_outline}. we denote these ground states by
\begin{align}
|\Omega ; 1\> = |+,+,...,+\> \nonumber \\
|\Omega ; s\> = |-,-,...,-\> 
\end{align}

The next step is to define domain wall states. Since the symmetry group is $\mathbb{Z}_2$, there is only one non-trivial domain wall state that we need to construct. Denoting this state by $|s_n\>$ where $n$ is the location of the domain wall, we define
\begin{align}
|s_n\>=|...,+, +, (+)_n, -, -, -, ...\>
\label{sndef}
\end{align}
Here we use the notation $(+)_n$ to indicate that the corresponding `$+$' is the state of the \emph{$n$th} spin, so that the `-' that follows is the state of the $(n+1)$st spin. Notice that (\ref{sndef}) implies a particular convention for labeling domain wall locations: a domain wall is at ``position $n$'' if the $n$th and $(n+1)$st spins are anti-aligned. (This will also hold for the multi-domain wall states.) Likewise, we define the trivial or ``no-domain'' wall state $|1_n\>$ in the obvious way:
\begin{align}
|1_n\>=|..., +, (+)_n, +, ...\>
\end{align}

Next we construct movement and splitting operators for these domain walls. We will start by constructing the movement operators. We define the movement operator between $n$ and $n+1$ by
\begin{align}
M^s_{(n+1)n}&=\sigma^+_{n+1}+U \sigma^+_{n+1}U^{-1} \nonumber \\
&=\sigma^+_{n+1}-\sigma^z_{n}\sigma^-_{n+1}\sigma^z_{n+2}
\end{align}
where the second equality follows from (\ref{z2lattice_sym}). Let us check that $M^s_{(n+1)n}$ obeys all the required conditions, i.e. $M^s_{(n+1)n}$ is local, $\mathbb{Z}_2$ symmetric, and has the correct action on domain walls. Locality and $\mathbb{Z}_2$ symmetry are obvious since $M^s_{(n+1)n}$ is explicitly symmetrized. As for the action on domain walls, this is easy to verify:
\begin{align}
M^s_{(n+1)n} |s_n\>&= \sigma^+_{n+1} |...,+, +, (+)_n, -, -, -, ...\> \nonumber \\
&= |...,+, +, (+)_n, +, -, -, ...\> \nonumber \\
&=|s_{n+1}\>
\end{align}
We define the reverse movement operator in a similar fashion:
\begin{align}
M^s_{n(n+1)}&=\left(M^s_{(n+1)n}\right)^\dagger \nonumber \\
&=\sigma^-_{n+1}-\sigma^z_{n}\sigma^+_{n+1}\sigma^z_{n+2}
\end{align}

As for the splitting operators, we define $S(s,s)$ as
\begin{align}
S(s,s)&=\sigma^-_{2}+U \sigma^-_{2}U^{-1} \nonumber \\
&=\sigma^-_2-\sigma^z_1\sigma^+_{2}\sigma^z_{3}
\end{align}
Again let us check that $S(s,s)$ is a valid splitting opreator. Clearly $S(s,s)$ is local and $\mathbb{Z}_2$ symmetric. To see that it has the correct action on domain walls note that
\begin{align}
S(s,s)|1_1\> &= \sigma^-_2|...,(+)_1, +, +,...\> \nonumber \\
&= |...,(+)_1, -, +,...\> \nonumber \\
&\propto |s_1, s_2\>
\end{align}
where $|s_1, s_2\>$ denotes the state with two domain walls at positons $1$ and $2$. 

Moving on to the other splitting operators, we define $S(1,s)$ as
\begin{align}
S(1,s)
&= M_{21}^s = \sigma^+_2-\sigma^z_1\sigma^-_{2}\sigma^z_{3}
\end{align}
Also, we define the movement operator $M^1_{n'n}$ and the splitting operators $S(s,1), S(1,1)$ as
\begin{align}
M^1_{n'n}=S(s,1)=S(1,1)=\mathbbm{1}
\end{align}
(The reason we can set these operators equal to the identity is that none of these operators are supposed to change the location of any nontrivial domain walls, e.g. $S(s,1)$ is defined by the condition $S(s,1)|s_1\> = |s_1, 1_2\>$, and similarly for the other operators).

With these operator definitions in place, we are now ready to calculate $F(s,s,s)$. Using Eq.~\ref{fdef1}, we have 
\begin{align}
|1\> &= M^s_{12} M^s_{01}S(s,s)M^s_{32}S(1,s)|s_1\> 
\end{align}
We now simplify this expression, working from right to left. First, we note that
\begin{align*}
S(1,s) |s_1\> &= (\sigma^+_2-\sigma^z_1\sigma^-_{2}\sigma^z_{3})|s_1\> \nonumber \\
&= \sigma^+_2|s_1\>
\end{align*}
since $\sigma^z_1\sigma^-_{2}\sigma^z_{3}|s_1\> = 0$. Likewise,
\begin{align*}
M^s_{32}\sigma^+_2|s_1\> = \sigma^+_3 \sigma^+_2 |s_1\>
\end{align*}
since the second term in $M^s_{32}$, i.e. $(-\sigma^z_{2}\sigma^-_{3}\sigma^z_{4})$, annihilates $\sigma^+_2|s_1\>$. Proceeding in this way, we can drop either the first or second term in each of the movement and splitting operators. The final result is:
\begin{align}
|1\> &=(-\sigma^z_1\sigma^+_2\sigma^z_3) \sigma^-_1 \sigma^-_2 \sigma^+_3 \sigma^+_2 |s_1\> \nonumber \\
&= |...,+,(-)_1,+,+,-,...\>
\end{align}
where the second equality follows from $|s_1\> = |...,+,(+)_1,-,-,-,...\>$. 
Following the same logic, we obtain
\begin{align}
|2\> &=M^s_{32}S(s,s)M^{1}_{12}M^s_{01}S(s,1)|s_1\> \nonumber\\
&=\sigma^+_3 (-\sigma^z_1\sigma^+_2\sigma^z_3) \sigma^-_1 |s_1\> \nonumber \\
&= -|...,+,(-)_1,+,+,-,...\>
\end{align}

Comparing these two expressions, we see that $|1\> = -|2\>$, so that
\beq
F(s,s,s)=\<2|1\>=-1
\eeq
More generally, one can check that all other values of $F$ are $1$ for this model, i.e. $F(g,h,k) = 1$
for all other choices of $g,h,k\in \{1,s\}$. 

Having computed $F$, the next question is to determine whether $F$ corresponds to a trivial or non-trivial cocycle, i.e. a trivial or non-trivial element of $H^3(G, U(1)) = \mathbb{Z}_2$. To answer this question, we compute the following gauge invariant quantity:
\begin{align}
F(s,s,s) F(s,1,s) = -1
\end{align}
Since this quantity is different from $1$, it follows that $F$ is a nontrivial cocycle. We conclude that our edge theory describes the boundary of the non-trivial bosonic SPT phase with $\mathbb{Z}_2$ symmetry. This conclusion is consistent with the original microscopic derivation of this edge theory~\cite{ChenMatrix,LevinGu}.


\subsection{Chiral boson edge theory for $\mathbb{Z}_2$ SPT phase}
\label{sec:z2field}

We now present an example involving a \emph{continuum} edge theory for the $\mathbb{Z}_2$ bosonic SPT phase (the same SPT phase as in the previous example). This edge theory was introduced in Ref.~\onlinecite{LevinGu}.

\subsubsection{Edge theory}
We begin by reviewing the continuum $\mathbb{Z}_2$ SPT edge theory.  
This edge theory is a chiral boson edge theory consisting of two conjugate fields $\theta,\phi$ obeying the commutation relations
\begin{align}
[\theta(x),\partial_y\phi(y)] &=2\pi i \delta(x-y) 
\label{thetaphicomm}
\end{align}
with all other commutators vanishing.

Again, to define the edge theory, we need to specify three pieces of data: (1) a Hilbert space $\mathcal{H}$; (2) a set of local operators $\{\mathcal{O}\}$ that act in $\mathcal{H}$; and (3) a collection of symmetry transformations $\{U^g : g \in G\}$ acting within $\mathcal{H}$.
The Hilbert space $\mathcal{H}$ is the usual infinite dimensional representation of the above algebra (\ref{thetaphicomm}). The local operators $\{\mathcal{O}\}$ in this edge theory consist of arbitrary derivatives and/or products of the operators $\{e^{\pm i \theta}, e^{\pm i \phi}\}$. 

To complete the edge theory, we need to specify the $\mathbb{Z}_2$ symmetry transformation, $U \equiv U^s$, where $\mathbb{Z}_2 = \{1, s\}$. This transformation acts as~\cite{LevinGu}
\begin{align}
      U \theta U^{-1} &= \theta -\pi \nonumber \\
      U \phi U^{-1} &= \phi -\pi
\label{z2symmcont2}
\end{align}
We can see that this is a $\mathbb{Z}_2$ symmetry since the fields $\theta,\phi$ are only defined modulo $2\pi$.

We can also write out an explicit formula for $U$, though we will not need it for our computation below~\cite{LevinGu}:
\begin{align}
U = e^{-\frac{i}{2}\int_{-\infty}^\infty dy [\partial_y \theta(y) + \partial_y \phi(y)]}
\label{z2symmcont1}
\end{align}

\subsubsection{Calculating the anomaly}

We now proceed to calculate the anomaly in the above edge theory. The first step is to choose a gapped Hamiltonian that breaks the $\mathbb{Z}_2$ symmetry spontaneously and completely. We will use the Hamiltonian
\beq
H= H_0 -\int dx V \cos(2\theta)
\eeq
where $H_0$ is the usual free boson Hamiltonian:
\begin{equation*}
H_0=\int dx \frac{1}{4\pi} [v_\theta (\partial_x \theta)^2 + v_\phi (\partial_x \phi)^2]
\end{equation*}
for some velocities $v_\theta, v_\phi > 0$.

When $V$ is sufficiently large, the  above Hamiltonian $H$ has all the required properties. In that regime, the cosine term locks $\theta$ to one of two values: $\theta = 0$ and $\theta = \pi$, spontaneously breaking the $\mathbb{Z}_2$ symmetry (\ref{z2symmcont2}) and opening up an energy gap.

For simplicity, we will consider the limit $V \rightarrow \infty$ in what follows. In this limit, the two ground states of $H$ are \emph{eigenstates} of $e^{i \theta(x)}$. Following our standard labeling scheme, we denote these ground states by $|\Omega;1\>$ and $|\Omega;s\>$, where 
\begin{align}
e^{i\theta(x)}|\Omega;1\> &= |\Omega;1\> \nonumber \\
e^{i\theta(x)}|\Omega;s\> &= -  |\Omega;s\> 
\end{align}
 for all $x$.

The next step is to construct domain wall states $|s_x\>$, that interpolate spatially between the two ground states, $|\Omega; 1\>$ and $|\Omega; s\>$. We define $|s_x\>$ by
\begin{align}
|s_x\>=a_x^\dagger|\Omega;1\>
\label{sx_def}
\end{align}
where 
\beq
a^\dagger_x=e^{-\frac{i}{2}\int_{x}^\infty dy \ \partial_y \phi(y)}
\label{adagdef}
\eeq
Here $a^\dagger_x$ is a (non-local) creation operator for a domain wall at postion $x$. To see that $|s_x\>$ is a valid domain wall state, note that $|s_x\>$ obeys
\begin{align}
e^{i\theta(x')}|s_x\> &= \text{sgn}(x-x')|s_x\> 
\end{align}
(This follows from the fact that $a_x^\dagger$ anticommutes with $e^{i \theta(x')}$ for $x' > x$). Likewise, we define the no-domain wall state $|1_x\>$ to be $|1_x\> = |\Omega; 1\>$.

Now that we have defined domain wall states, the next step is to construct movement and splitting operators for these domain walls. First, we define the movement operator $M^s_{x'x}$ by
\beq
M^s_{x'x}=e^{\frac{i}{2}\int_x^{x'} dy \ \partial_{y}\phi(y)}
\eeq
This is a valid movement operator because it is local and $\mathbb{Z}_2$ symmetric and it obeys $M^s_{x'x}|s_x\>\propto|s_{x'}\>$ since $M^s_{x'x}a^\dagger_x\propto a^\dagger_{x'}$.

Next, we define the splitting operator $S(s,s)$ by
\begin{align}
S(s,s)&=M^s_{21}e^{i[\phi(1)-\theta(1^-)]} \nonumber \\
&=e^{\frac{i}{2}\int_1^{2} dy \ \partial_{y}\phi(y)} e^{i[\phi(1)-\theta(1^-)]}
\end{align}
Here, $\theta(1^-)$ is shorthand for $\theta(1-\epsilon)$ where $\epsilon$ is a small positive number.

To see that this is a valid splitting operator, note that $S(s,s)$ is local, it is $\mathbb{Z}_2$ symmetric, and furthermore
\begin{align}
S(s,s)|1_1\> 
&=  e^{\frac{i}{2}\int_1^{2} dy \ \partial_{y}\phi(y)} e^{i[\phi(1)-\theta(1^-)]} |\Omega; 1\> \nonumber \\
&\propto e^{\frac{i}{2}\int_1^{2} dy \ \partial_{y}\phi(y)} e^{i\phi(1))} |\Omega; 1\> \nonumber \\
&\propto a^\dagger_2 a^\dagger_1 |\Omega; 1\> \nonumber \\
&\propto |s_1, s_2\>
\end{align}
Here, the second equality follows from $e^{-i \theta(1^-)} |\Omega; 1\> = |\Omega; 1\>$, while the third equality follows from
the definition of $a^\dagger_x$ (\ref{adagdef}). Readers may wonder why we include the factor of $e^{-i \theta(1^-)}$ in the definition $S(s,s)$ given that $e^{-i \theta(1^-)}$ acts trivially on $|\Omega; 1\>$: the reason for including this factor of $e^{-i \theta(1^-)}$ is that, without it, $S(s,s)$ would be \emph{odd}, not even, under the $\mathbb{Z}_2$ symmetry. Furthermore, the reason that we use $e^{-i \theta(1^-)}$ rather than say, $e^{-i \theta(1)}$ is that this choice will regularize some of the commutators that we calculate below.  (We could equally well choose $e^{-i \theta(1^+)}$ and we would arrive at the same result.)

Moving on to the other splitting operators, we define
\begin{align}
S(1,s)=M^s_{21}
\end{align}
Also we define
\begin{align}
M^1_{x'x}=S(s,1)=S(1,1)=1
\end{align}

With these operators in hand, we are now ready to compute $F(s,s,s)$. Using (\ref{fdef1}), we have
\begin{align}
|1\> &= M^s_{12} M^s_{01}M^s_{21}e^{i[\phi(1)-\theta(1^-)]}M^s_{32}M^s_{21}|s_1\> \nonumber \\
|2\> &=M^s_{32}M^s_{21}e^{i[\phi(1)-\theta(1^-)]}M^s_{01}|s_1\> 
\end{align}

In order to compare these expressions, we need to reorder the operators within them. We do this using the following commutation relations, which can be derived from the Baker-Campbell-Hausdorff formula:
\begin{align}
e^{i[\phi(1)-\theta(1^-)]} M^s_{x1} &= (-1)^{\Theta(1-x)} M^s_{x1} e^{i[\phi(1)-\theta(1^-)]} \nonumber \\
[M^s_{x'x}, M^s_{y'y}]  &= 0 
\end{align}
where $\Theta(x)$ denotes the Heaviside step function. With these formulas and the identity $M^s_{xx'}=(M^s_{x'x})^{-1}$, we can rewrite $|1\>$ as:
\begin{align}
|1\> &=-M^s_{32}M^s_{21}e^{i[\phi(1)-\theta(1^-)]}M^s_{01}|s_1\> 
\end{align}
Therefore, $|1\> = -|2\>$ and hence
\begin{align}
F(s,s,s)=\<2|1\> =-1
\end{align}
In the same way, one can check that all other values of $F$ are $1$ for this model, i.e. $F(g,h,k) = 1$, for all other choices of $g,h,k \in \{1,s\}$. 

Comparing with the previous example, we see that the two $F$'s are identical. Therefore, just as in that example, we conclude that $F$ is a non-trivial cocycle, and the correponding edge theory describes the boundary of the non-trivial $\mathbb{Z}_2$ SPT phase. This result is consistent with the original derivation of this edge theory~\cite{LevinGu}.


\subsection{SPT lattice edge theory with symmetry group $G$}
\label{sec:general_lattice}

We now generalize the example in Sec.~\ref{sec:z2lattice} to a large class of lattice edge theories with a finite unitary symmetry group $G$. A similar class of edge theories was studied in Ref.~\onlinecite{Else} and we will mostly follow their notation here.

\subsubsection{Edge theory}

As before, to define the edge theory, we need to specify three pieces of data: (1) a Hilbert space $\mathcal{H}$; (2) a set of local operators $\{\mathcal{O}\}$ that act in $\mathcal{H}$; and (3) a collection of symmetry transformations $\{U^g : g \in G\}$ acting within $\mathcal{H}$.

We begin by describing the Hilbert space $\mathcal{H}$. This Hilbert space is equivalent to a one-dimensional spin chain where each spin can be in $|G|$ different states. We label these states by group elements, $|g\>$, $g \in G$. In this notation, the basis states for the Hilbert space are of the form
\begin{align}
|...,\alpha_{-1},\alpha_0,\alpha_1,...\>, \quad \quad \alpha_n\in G
\end{align}
The local operators $\{\mathcal{O}\}$ in this edge theory are the usual local operators in a spin chain -- i.e. products of single site operators acting on a collection of neighboring sites. In particular, there are two basic types of single site operators, from which all other operators can be built. The first operator, $P^g_n$, is a projection operator that projects onto states with $\alpha_n = g$, i.e.
\begin{align}
P^g_n |...,\alpha_{n-1},\alpha_n,\alpha_{n+1},...\> = \delta_{\alpha_n, g}|...,\alpha_{n-1},g,\alpha_{n+1},...\>
\label{pgn_def}
\end{align}
The second operator, $R^g_n$ is a unitary operator that performs right multiplication by $g$ on the $n$th spin:
\begin{align}
R^g_n |...,\alpha_{n-1},\alpha_n,\alpha_{n+1},...\> = |...,\alpha_{n-1},\alpha_n g,\alpha_{n+1},...\>
\end{align}

To complete the edge theory, we need to describe the symmetry transformations $U^g$. We start by writing down an explicit formula for $U^g$. This formula is a product of two terms:
\begin{align}
U^g=N^g X^g
\end{align}
where $X^g$ acts by left multiplication by $g$ on every site,
\begin{align}
X^g|...,\alpha_{-1},\alpha_0,\alpha_1,...\>=|...,g\alpha_{-1},g\alpha_0,g\alpha_1,...\>,
\end{align}
while $N^g$ is a phase factor that is diagonal in the $|\alpha\>$ basis:
\begin{align}
N^g|\alpha\>= e^{i\mathcal{N}^{(1)}(g)[\alpha]}|\alpha\>
\end{align}
Here we are using the abbreviation $|\alpha\> \equiv |...,\alpha_{-1},\alpha_0,\alpha_1,...\>$. The term in the exponent, $\mathcal{N}^{(1)}(g)$, is a functional of the configuration $\alpha$ of the form
\begin{align}
\mathcal{N}^{(1)}(g)[\alpha] = \sum_{n} \Phi_n^g(\alpha_{n-1}, \alpha_n, \alpha_{n+1})
\end{align}  
for some real valued $\Phi_n^g$ that depends on \emph{triplets} of neighboring spins. We note that the above functional form is not essential for our method -- we could equally well consider generalizations of $\Phi_n^g$ that act on any finite number of nearby spins.

An alternative, and more local, way to describe the symmetry transformations $U^g$ is to specify how these transformations act on local operators -- specifically $P_n^h$ and $R_n^h$. In this description, the symmetry transformation is defined by
\begin{align}
U^g P^h_n (U^g)^{-1}  &= P^{g h}_n \nonumber \\
U^g R^{h}_n (U^g)^{-1} &=R^h_nW^{g,h}_n\label{eq:R_trans}
\end{align}
where $W^{g,h}_n$ is an unitary operator of the form
\begin{align}
W^{g,h}_n = \sum_{a,b,c \in G} \Theta_n^{g,h}(a,b,c) P^{a}_{n-1} P^{b}_n P^{c}_{n+1}\label{eq:W_def}
\end{align}
and where $\Theta_n^{g,h}(a,b,c)$ is a $U(1)$ phase. (Again, our method does not require this particular functional form of $\Theta_n^{g,h}$ and we could consider generalizations that act on any finite number of nearby spins). In the calculation that follows, we only use the latter, more local, description of the symmetry transformation given in Eqs.~\ref{eq:R_trans}-\ref{eq:W_def}.

\subsubsection{Calculating the anomaly}

The first step is to choose an edge Hamiltonian that breaks the $G$-symmetry spontaneously and completely. We choose
\begin{align}
H = -J \sum_n \sum_g P^g_n P^g_{n+1}
\end{align} 
Note that $H$ has a ``ferromagnetic'' interaction that favors states in which neighboring spins are in the same state $|g\>$.
As a result, it is easy to see that this Hamiltonian has $|G|$ degenerate ground states of the form $|... , g, g, g, ...\>$ where $g \in G$. We will label these states by
\begin{align}
|\Omega; g\> = |..., g, g, g, ...\>
\end{align}

We now turn to the definition of the domain wall state $|g_n\>$. We define $|g_n\>$ by
\begin{align}
|g_n\> = |..., 1, 1, (1)_n, g, g, g, ...\>
\end{align}
Here the notation $(1)_n$ signifies that this `1' is the state of the $n$th spin, so that the `g' that follows is the state of the $(n+1)$st spin and so on. Similarly to Sec.~\ref{sec:z2lattice}, the domain wall at location $n$ sits between sites $n$ and $n+1$.

Having defined the domain wall states, the next step is to construct movement and splitting operators. We define the movement operator between $n$ and $n+1$ by
\begin{align}
M^g_{n+1,n}&=\sum_{k\in G} U^k R^{g^{-1}}_{n+1}P^g_{n+1}(U^k)^{-1}
\end{align}
To see that $M^g_{n+1,n}$ is a valid movement operator notice that it is local and $G$-symmetric by construction.
Furthermore, it has the correct action on domain wall states:
\begin{align}
M^g_{(n+1)n}|g_n\>&=\sum_{k\in G} U^k R^{g^{-1}}_{n+1}P^g_{n+1}(U^k)^{-1} |..., (1)_n,g, g,...\> \nonumber \\
&=R^{g^{-1}}_{n+1}|..., (1)_n,g, g,...\> \nonumber \\
&= |..., (1)_n,1, g,...\> \nonumber \\
&= |g_{n+1}\>
\end{align}
Here the second equality follows from noting that all the terms in the sum vanish except for $k=1$. Similarly, we define the reverse movement operator by
\begin{align}
M^g_{n(n+1)}=\left(M^g_{(n+1)n}\right)^\dagger
\end{align}
Following a similar calculation to before, one can check that this is a valid movement operator.

Moving on to splitting operators, we define $S(g,h)$ by
\begin{align}
S(g,h)&=\sum_{k\in G} U^k R^{h^{-1}}_{2}P^{gh}_{2}(U^k)^{-1}
\end{align}
Again, $S(g,h)$ is local and $G$-symmetric by construction. We now show that it has the correct action on domain wall states:
\begin{align}
S(g,h)|gh_1\>&=\sum_{k\in G} U^k R^{h^{-1}}_{2}P^{gh}_{2}(U^k)^{-1}|..., (1)_1, gh, gh,...\> \nonumber \\
&=R^{h^{-1}}_{2} |..., (1)_1, gh, gh,...\> \nonumber \\
&= |..., (1)_1, g, gh,...\> \nonumber \\
&= |g_1,h_2\>
\end{align}
Again, the second equality follows from noting that all the terms in the sum vanish except for $k=1$.

We are now ready to calculate $F(g,h,k)$. Using (\ref{fdef1}), we have
\begin{align}
|1\> =& M^h_{12} M^g_{01} S(g,h) M^k_{32} S(gh,k) |ghk_1\> \nonumber \\
=&(U^g R^{h}_2 [U^g]^{-1}) \cdot R^g_1 \cdot R^{h^{-1}}_2 \nonumber \\
&\cdot (U^{gh} R^{k^{-1}}_3 [U^{gh}]^{-1}) \cdot R^{k^{-1}}_2 |ghk_1\> 
\label{1genG}
\end{align}
Here, to derive the second equality, notice that whenever a movement or splitting operator acts on a domain wall state, only one value of $k$ gives a nonzero contribution; Eq.~\ref{1genG} follows by keeping this one nonvanishing term for each movement and splitting operator.

To proceed further, we use the transformation law for $R^h_n$ in Eqs.~\ref{eq:R_trans}-\ref{eq:W_def} to derive
\begin{align}
|1\> =& (R^h_2 W_2^{g,h}) \cdot R^g_1 \cdot R^{h^{-1}}_2 \cdot (R^{k^{-1}}_3 W_3^{gh,k^{-1}}) \cdot R^{k^{-1}}_2 |ghk_1\> \nonumber \\
=& \Theta_2^{g,h}(g,g,gh) \cdot \Theta_3^{gh,k^{-1}}(gh,ghk,ghk) \nonumber \\
&\cdot |...,1,(g)_1,gh,gh,ghk,...\>
\end{align}
where the second equality follows from $|ghk_1\> = |...,1,(1)_1,ghk,ghk,ghk,...\>$.
By the same reasoning, we have
\begin{align}
|2\> =& M^k_{32} S(h,k) M^{hk}_{12} M^g_{01} S(g,hk) |ghk_1\> \nonumber \\
=& (U^{gh} R^{k^{-1}}_3 [U^{gh}]^{-1}) \cdot (U^g R^{k^{-1}}_2 [U^g]^{-1}) \nonumber \\
& \cdot (U^g R^{hk}_2 [U^g]^{-1}) \cdot R^{g}_1 \cdot R^{(hk)^{-1}}_2 |ghk_1\> \nonumber \\
=&(R^{k^{-1}}_3 W_3^{gh,k^{-1}}) \cdot (R^{h}_2W_2^{g,h}) \cdot R^{g}_1 \cdot R^{(hk)^{-1}}_2 |ghk_1\> \nonumber \\
=& \Theta_3^{gh,k^{-1}}(gh,ghk, ghk) \cdot \Theta_2^{g,h}(g,g,ghk) \nonumber \\
&\cdot |...,1,(g)_1,gh,gh,ghk,...\>
\end{align}

Taking the inner product between $|1\>$ and $|2\>$, we see that the $\Theta_3^{gh,k^{-1}}(gh,ghk, ghk)$ factors cancel out, leaving
\begin{align}
F(g,h,k)=\<2|1\> =\frac{\Theta^{g,h}_2(g,g,gh)}{\Theta^{g,h}_2(g,g,ghk)}
\end{align}
where we are using the fact that $\Theta_n^{g,h}$ is a $U(1)$ phase.

\section{Connection with symmetry restriction method for computing anomalies}
\label{sec:conn_symm_rest}

As we mentioned earlier, in Ref.~\onlinecite{Else}, Else and Nayak showed how to compute anomalies in a large class of SPT edge theories using restricted symmetry operators. It is natural to wonder how our $F$-symbol based approach is related to this symmetry restriction approach. In this section we derive a connection between the two approaches by explicitly showing that the two approaches give identical results in cases where both methods are applicable. 

\subsection{Review of symmetry restriction method}

We begin by reviewing the symmetry restriction method~\cite{Else}. This method applies to SPT edge theories with a discrete unitary symmetry group $G$ and with the property that the symmetry operators $\{U^g, g \in G\}$ are local unitary transformations.\footnote{This method also applies to continuous symmetries and some antiunitary symmetries; we focus on discrete unitary symmetries for simplicity.} Here, by a ``local unitary transformation'', we mean that $U^g$ can be generated by the time evolution of a local Hermitian operator over a finite period of time $T$: $U^g = \mathcal{T} \exp[ -i\int_0^T dt H(t)]$. 

The symmetry restriction method proceeds as follows. Consider an edge theory of the above kind, with symmetry operators $U^g$. To compute the anomaly associated with this edge theory, the first step is to choose a large interval $I = [a,b]$, and then choose a ``restriction'' of $U^g$ to $I$, which we will denote by $U_{I}^g$. Here, when we say that $U_{I}^g$ is a restriction of $U^g$ to $I$, we mean $U_{I}^g$ has two properties: (i) $U_I^g$ is a local unitary transformation supported in a neighborhood of $I$, and (ii) for any operator $\mathcal{O}$ that is supported in $I$, 
\begin{equation}
U_I^g \mathcal{O} (U_I^g)^{-1} =U^g \mathcal{O}(U^g)^{-1}
\label{restrict_def}
\end{equation}
Note that the existence of such a $U_I^g$ is guaranteed by the fact that $U^g$ is a local unitary transformation, but $U_I^g$ is not unique.

Next, define an operator $\Omega_I(g,h)$ by
\begin{align}
\Omega_I(g,h) = U_I^g U_I^h (U_I^{gh})^{-1}
\label{omegaidef}
\end{align}
By construction $\Omega_I(g,h)$ is a local unitary transformation that is supported near $a,b$ -- the endpoints of $I$. It follows that we can factor $\Omega_I(g,h)$ as a product
\begin{align}
\Omega_I(g,h)=\Omega_a(g,h)\Omega_b(g,h)
\label{omegaifact}
\end{align}
where $\Omega_a(g,h)$ and $\Omega_b(g,h)$ are unitary operators supported near $a$ and $b$, respectively.\footnote{Readers may notice that there is a phase ambiguity in $\Omega_a(g,h)$, $\Omega_b(g,h)$, i.e. we can replace $\Omega_a(g,h) \rightarrow \Omega_a(g,h) \nu(g,h)$, and $\Omega_b(g,h) \rightarrow \Omega_b(G,h) \nu^{-1}(g,h)$ where $\nu(g,h)$ is a $U(1)$ phase. This ambiguity is related to the fact that the quantity $\omega(g,h,k)$ is only well-defined up to a coboundary.}

The operator $\Omega_a(g,h)$ (or equivalently $\Omega_b(g,h)$) is the key to computing the anomaly. In particular, Ref.~\onlinecite{Else} showed that $\Omega_a$ and $\Omega_b$ obey the following operator identities:
\begin{align}\label{elseomega}
\Omega_a(g,h)\Omega_a(gh,k)&=\omega(g,h,k)U_I^g\Omega_a(h,k)(U_I^g)^{-1}\Omega_a(g,hk) \nonumber \\
\Omega_b(g,h)\Omega_b(gh,k)&=\nonumber\\
\omega^{-1}(g,&h,k)U_I^g\Omega_b(h,k)(U_I^g)^{-1}\Omega_b(g,hk)
\end{align}
where $\omega(g,h,k) \in H^3(G, U(1))$ is the anomaly carried by the SPT edge theory. Thus, if we know $\Omega_a(g,h)$ (or equivalently $\Omega_b(g,h)$), we can immediately compute the anomaly by comparing the left and right hand sides of Eq.~\ref{elseomega}.

Putting this all together, the symmetry restriction method involves the following steps: one first computes the restricted symmetry operator $U_I^g$, and then the associated operators $\Omega_I(g,h)$ and $\Omega_a(g,h)$. One then computes the anomaly $\omega(g,h,k)$ using Eq.~\ref{elseomega} above.

\subsection{$F$-symbol computation}

We now show how to compute the $F$-symbol for the above class of edge theories, i.e. edge theories with a finite unitary symmetry group $G$ and with the property that the symmetry operators $\{U^g, g \in G\}$ are local unitary transformations. Our goal will be to show that $F(g,h,k) = \omega(g,h,k)$. 

Consider any edge theory of the above type. To compute the $F$-symbol for such an edge theory, the first step is to choose an edge Hamiltonian that breaks the symmetry spontaneously and completely and opens up a gap. We then label the $|G|$ degenerate ground states by $\{|\Omega ; g\>\}$ where
\begin{align}
|\Omega; g\> = U^g |\Omega ; 1\>
\end{align}

\begin{figure}[tb]
\centering
\includegraphics[width=1.0\columnwidth]{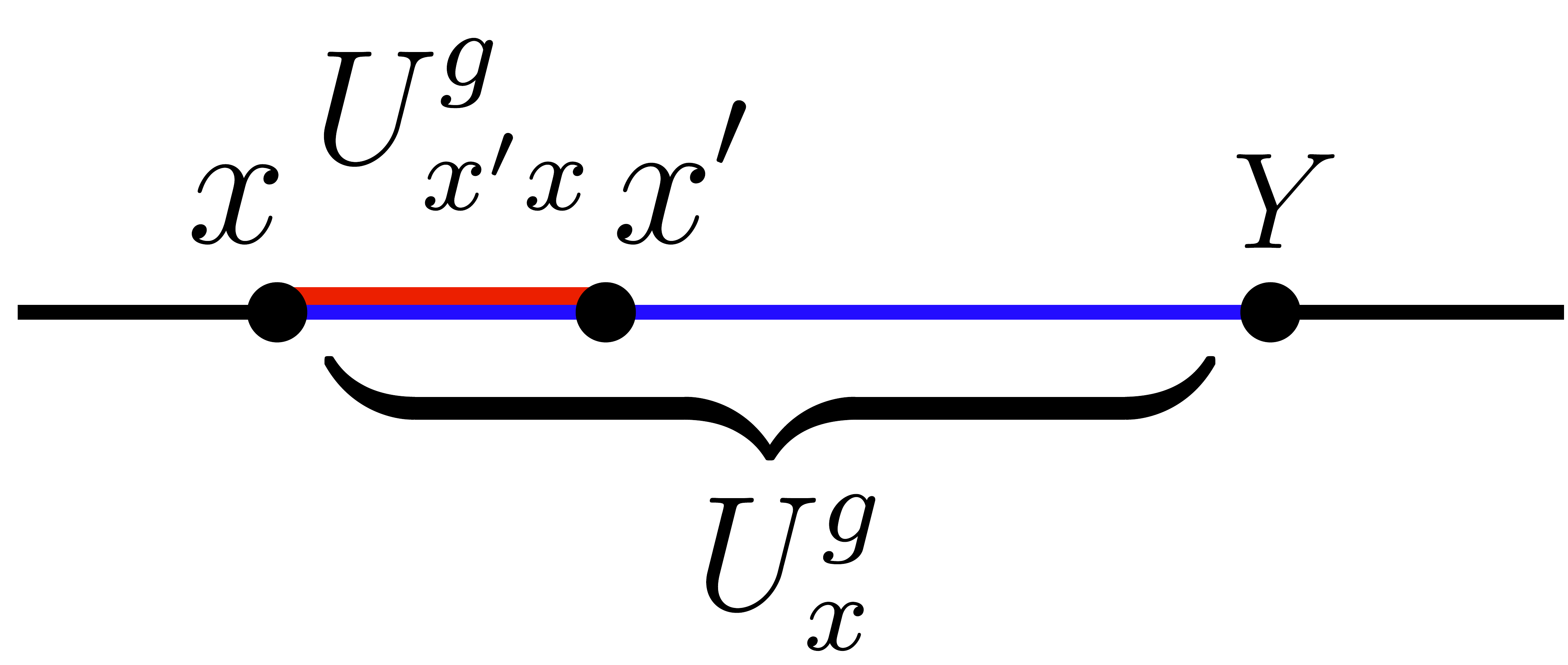}
\caption{Regions of support for the operators $U_{x'x}^g$ and $U_x^g$: the operator $U_x^g$ is supported in the interval $[x,Y]$ shown in blue, while $U_{x'x}^g$ is supported in the interval $[x,x']$ shown in red.}
\label{fig:Ux}
\end{figure}

The next step is to define domain wall states. We will do this using restricted symmetry operators, in order to facilitate a comparison with Ref.~\onlinecite{Else}. To begin, we choose a point $Y$ that is far to the right of the region where we will be manipulating domain walls. This point $Y$ can be thought of as playing a similar role to $+\infty$, but it will be important for our purposes that $Y$ is finite. Next, for every point $x < Y$, we choose a unitary operator $U_{x}^g$ that is a restriction of $U^g$ to the interval $[x,Y]$, where this restriction is defined as in Eq.~\ref{restrict_def} above. In addition, we require that the $U_x^g$ are chosen so that they obey the following matching condition: for any two points $x < x'$ the operators $U_x^g$ and $U_{x'}^g$ have the same action except in the neighborhood of the interval $[x,x']$. Equivalently, we will require that the operator
\begin{align}
U_{x'x}^g \equiv U_{x'}^g(U_{x}^g)^{-1}
\label{consassump}
\end{align}
is supported in a neighborhood of the interval $[x,x']$ (see Fig.~\ref{fig:Ux}). One can show that it is always possible to choose $U_x^g$ in this way, given our assumption that $U^g$ are local unitary transformations.

Having defined $U_x^g$, we now define (single) domain wall states $|g_x\>$ by
\begin{align}
|g_x\>=U_x^g|\Omega;1\>
\end{align}
By construction $U_x^g|\Omega;1\>$ contains a type $g$ domain wall at $x$ and a type $g^{-1}$ domain wall at $Y$, so one could view $U_x^g|\Omega;1\>$ as a \emph{two} domain wall state. However, we will view $U_x^g|\Omega;1\>$ as a single domain wall state. We can do this because (i) the point $Y$ is far away from the region of interest, and (ii) the $g^{-1}$ domain wall at $Y$ takes the same form independent of $x$: for any $x, x'$ the two states $U_x^g|\Omega;1\>$ and $U_{x'}^g|\Omega;1\>$ have the same local expectation values near $Y$, due to our assumption (\ref{consassump}). Actually, the above definition of $|g_x\>$ is no different than the one given in e.g. Eq.~\ref{sx_def} where we set $Y = +\infty$, except that we are now calling attention to the point $Y$ because it will be useful for proving that our formalism matches that of Else and Nayak.

Now that we have defined domain wall states, the next step is to construct splitting and movement operators. To this end, notice that the operator $U_{x'x}^g$ defined in Eq.~(\ref{consassump}) obeys 
\begin{align}
U_{x'x}^g|g_x\> = |g_{x'}\>
\end{align}
Given this observation, we can construct a movement operator $M^g_{x'x}$ by appropriately symmetrizing $U^g_{x'x}$:
\begin{align}
M^g_{x'x}=\sum_{k\in G} U^k U_{x'x}^g P^g_{x'} (U^k)^{-1}, \quad \quad x' > x
\end{align}
Here $P^g_{x'}$ is a projection operator that projects onto states that share the same local expectation values as $|\Omega; g\>$ in the neighborhood of $x'$. More precisely, $P^g_{x'}$ is a Hermitian operator that is supported in the neighborhood of $x'$ and that has the property that $P^g_{x'} |\Omega; h\> = \delta_{gh} |\Omega; g\>$. For the models discussed in Sec.~\ref{sec:general_lattice}, $P^g_{x'}$ can be defined is the same way as the operator $P^g_n$ (\ref{pgn_def}). In more general systems, it may be necessary to take the neighborhood around $x'$ to be of order $\xi$ to construct a projector $P^g_{x'}$ of this kind. In any case, we will take the existence of this projection operator as an assumption.

To see that $M^g_{x'x}$  is a valid movement operator, note that it is local and $G$-symmetric by construction. Furthermore, it has the correct action on domain walls:
\begin{align}
M^g_{x'x}|g_x\> =  U_{x'x}^g |g_x\> = |g_{x'}\>
\end{align}
Similarly, we define the reverse movement operator by 
\begin{align}
M^g_{x x'}= (M^g_{x'x})^\dagger, \quad \quad x' > x
\end{align}

In order to construct splitting operators, we first define a unitary operator $\Omega(g,h)$ by
\begin{align}
\Omega(g,h)=U_1^g U_1^h (U_1^{gh})^{-1}
\label{omegadef}
\end{align} 
By construction $\Omega(g,h)$ is a local unitary operator that is supported near $1$ and $Y$, so we can factor $\Omega(g,h)$ as
\begin{align}
\Omega(g,h)=\Omega_1(g,h)\Omega_Y(g,h)
\label{omegafact}
\end{align}
where $\Omega_1(g,h)$ and $\Omega_Y(g,h)$ are unitary operators supported near $1$ and $Y$, respectively. (Note that Eqs.~\ref{omegadef}-\ref{omegafact} are essentially the same as Eqs.~\ref{omegaidef}-\ref{omegaifact}, with $I = [1,Y]$). 

With this notation, we define the splitting operator $S(g,h)$ by
\begin{align}
S(g,h)=\sum_{k\in G} U^k U_1^g U_{21}^h (U_1^g)^{-1} \Omega_1(g,h)P^{gh}_{2} (U^k)^{-1}.
\end{align}
Again, it is easy to see that $S(g,h)$ is $G$-symmetric and supported in a neighborhood around $[1,2]$. To see that it has the correct action on domain wall states, note that
\begin{align}
S(g,h)|gh_1\>&=U_1^g U_{21}^h (U_1^g)^{-1}  \Omega_1(g,h)|gh_1\> \nonumber \\
&=\Omega_Y(g,h)^{-1} U_1^g U_{21}^h (U_1^g)^{-1}  \Omega(g,h)|gh_1\> \nonumber \\
&=\Omega_Y(g,h)^{-1} U_1^g U_2^h|\Omega;1\> \nonumber \\
&\propto |g_1,h_2\>.
\end{align}
Here, the last equality follows from observing that the state $\Omega_Y(g,h)^{-1} U_1^g U_2^h|\Omega;1\>$ has the same local expectation values as $|g_1\>$ near $x=1$ and the same local expectation values as $U^g |h_2\>$ near $x=2$, and the same local expectation values as $U^g |\Omega; 1\>$ in the region between $x=1$ and $x=2$. (We don't have to worry about the expectaton values away from $[1,2]$ since the locality of $S(g,h)$ guarantees that $S(g,h)|gh_1\>$ has the correct expectation values away from $[1,2]$).

Now that we have defined movement and splitting operators, we can compute the $F$-symbol using Eqs.~\ref{fdef1}-\ref{fdef2}, i.e.
\begin{align}
F(g,h,k) = \<2|1\>, 
\label{Fdefen}
\end{align}
where
\begin{align}
|1\>&= M^h_{12} M^g_{01} S(g,h) M^k_{32} S(gh,k) |ghk_1\> \nonumber \\
|2\>&= M^k_{32} S(h,k) M^{hk}_{12} M^g_{01} S(g,hk) |ghk_1\> 
\label{12def}
\end{align}

\subsection{Comparing results with Else and Nayak}

\begin{figure}[tb]
\centering
\includegraphics[width=1.0\columnwidth]{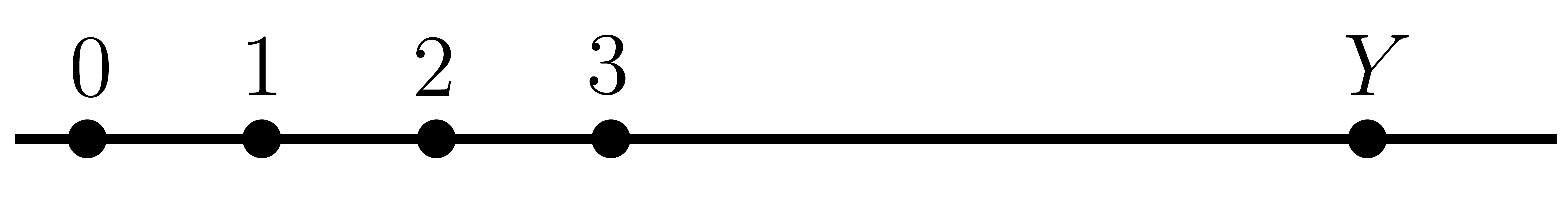}
\caption{The points that are relevant to the calculation of $F(g,h,k)$ and $\omega(g,h,k)$: the operators that are used to calculate $F(g,h,k)$ are supported in a neighborhood of $[0,3]$, whereas the operators used to calculate $\omega(g,h,k)$ are supported in a small neighborhood around the point $Y$, far to the right of $3$. We connect the two calculations using the identity $\Omega_1(g,h) = \Omega_Y^{-1}(g,h) \Omega(g,h)$.} 
\label{fig:comparison}
\end{figure}

We now show $F(g,h,k) = \omega(g,h,k)$ where $F(g,h,k)$ is defined in Eq.~\ref{Fdefen} and $\omega(g,h,k)$ is defined in Eq.~\ref{elseomega}.

To prove this equality, we use the fact that $\Omega_1(g,h) = \Omega_Y^{-1}(g,h) \Omega(g,h)$. Substituting this identity into the definition of $S(g,h)$ and simplifying gives
\begin{align}
S(g,h) = \sum_{k\in G} U^k \Omega_Y^{-1}(g,h) U_1^g U_{2}^h (U_1^{gh})^{-1} P^{gh}_{2} (U^k)^{-1}
\label{spliten}
\end{align}
The basic idea of the proof is to substitute (\ref{spliten}) into (\ref{12def}) and then simplify the resulting expressions for $|1\>, |2\>$. After this substitution, we will then separate out all the $\Omega_Y$ factors. These $\Omega_Y$ factors will naturally form combinations like those in Eq.~\ref{elseomega}, which will allow us to derive a direct connection between the F-symbol $F(g,h,k)$ and the cocycle $\omega(g,h,k)$ (see Fig.~\ref{fig:comparison}).

Following this plan, we first simplify $|1\>$ as follows:
\begin{align}
|1\>&=M^{h}_{12}M^{g}_{01}S(g,h)M^{k}_{32} S(gh,k)|ghk_1\> \nonumber \\
&= M^{h}_{12}M^{g}_{01}S(g,h)M^{k}_{32} \Omega_Y^{-1}(gh,k) U_1^{gh} U_{2}^k (U_1^{ghk})^{-1} |ghk_1\> \nonumber \\
&=\Omega_Y^{-1}(gh,k) M^{h}_{12}M^{g}_{01}S(g,h)M^{k}_{32}U_1^{gh}U_2^k|\Omega;1\>\nonumber \\
&= \Omega_Y^{-1}(gh,k) M^{h}_{12}M^{g}_{01}S(g,h)U_1^{gh}U_3^k|\Omega;1\>\nonumber \\
&=[\Omega_Y^{-1}(gh,k) \Omega_Y^{-1}(g,h)] M^{h}_{12}M^{g}_{01}U_1^g U_2^h U_3^k|\Omega;1\>\nonumber \\
&=[\Omega_Y^{-1}(gh,k) \Omega_Y^{-1}(g,h)] U_{0}^g U_1^h U_3^k|\Omega;1\>
\label{1en}
\end{align}
Here, the second and fifth equalities follow from substituting (\ref{spliten}), while the third equality follows from the fact that $\Omega_Y(gh,k)$ is supported near $Y$ and therefore commutes with all the movement and splitting operators. Likewise, the fourth and sixth equalities follow from the fact that $[M_{x'x}^g, U^h_y] = 0$ as long as $y \notin [x, x']$, which in turn follows from the fact that the movement operators are $G$-symmetric.

We can simplify $|2\>$ in a similar manner:
\begin{align}
|2\>&=M^{k}_{32}S(h,k)M^{hk}_{12}M^{g}_{01}S(g,hk)|ghk_1\>\nonumber\\
&= \Omega_Y^{-1}(g,hk) M^{k}_{32}S(h,k)M^{hk}_{12}M^{g}_{01}U_1^g U_2^{hk}|\Omega;1\>\nonumber\\
&=\Omega_Y^{-1}(g,hk) M^{k}_{32}S(h,k)U_0^g U_1^{hk}|\Omega;1\>\nonumber\\
&=\Omega_Y^{-1}(g,hk) M^{k}_{32}U_0^g S(h,k)U_1^{hk}|\Omega;1\>\nonumber\\
&= [\Omega_Y^{-1}(g,hk) U_0^g \Omega_Y^{-1}(h,k) (U_0^g)^{-1}] M^{k}_{32}U_0^g U_1^h U_2^k|\Omega;1\>\nonumber\\
&=[\Omega_Y^{-1}(g,hk) U_0^g \Omega_Y^{-1}(h,k) (U_0^g)^{-1}] U_0^g U_1^h U_3^k|\Omega;1\> \nonumber \\
&=[\Omega_Y^{-1}(g,hk) U_1^g \Omega_Y^{-1}(h,k) (U_1^g)^{-1}] U_0^g U_1^h U_3^k|\Omega;1\>
\label{2en}
\end{align}
where the last line is obtained the fact that $\Omega_Y^{-1}(h,k)$ is supported near $Y$. To make contact with (\ref{elseomega}), notice that the inverse of the second equation in (\ref{elseomega}) with $I = [1,Y]$ gives the following operator identity:
\begin{align}
\Omega_Y^{-1}(gh,k) &\Omega_Y^{-1}(g,h) = \nonumber \\
&\omega(g,h,k) \Omega_Y^{-1}(g,hk) U_1^g\Omega_Y^{-1}(h,k)(U_1^g)^{-1}
\end{align}
Comparing this identity with the bracketed expressions in Eqs. \ref{1en} and \ref{2en}, we conclude that
\begin{align}
|1\> = \omega(g,h,k) |2\>
\end{align}
Hence,
\begin{align}
F(g,h,k) = \<2| 1\> = \omega(g,h,k),
\end{align}
as we wished to show.


\section{Discrete antiunitary symmetries}
\label{sec:antiunitary}

So far we have focused on the case where the symmetry group $G$ is \emph{unitary}. In this section, we consider the more general case where $G$ contains both unitary and antiunitary symmetry transformations.

\subsection{Cohomology group}
\label{sec:antiunitary_cohom}
We start by reviewing the cohomology group $H^3(G, U_T(1))$. This group is important because it describes the output of our procedure in the general antiunitary case. 

First, we explain the meaning of ``$U_T(1)$.'' This symbol denotes the group $U(1)$ with a particular $G$-module structure, which is defined as follows: for any $g \in G$ and $\omega = e^{i\theta} \in U(1)$, the action of $g$ on $\omega$ is given by 
\begin{align}
    g \omega = \begin{cases} \omega^* & \text{ $g$ antiunitary} \\
    \omega & \text{ $g$ unitary}
    \end{cases}
\label{gaction}
\end{align}
In other words, antiunitary symmetries act on $U(1)$ by complex conjugation, while unitary symmetries act trivially. 

Just like the unitary case, the cohomology group $H^3(G, U_T(1))$ consists of equivalence classes of functions $\omega:G\times G\times G\rightarrow U(1)$ obeying a ``cocycle'' condition. However, the cocycle condition takes a modified form, namely
\begin{align}
\frac{\omega(g,h,k) \omega(g,hk,l) [g\omega(h,k,l)]}{\omega(gh,k,l) \omega(g,h,kl)} = 1
\label{cocycleanti}
\end{align}
where $g\omega$ denotes the group action (\ref{gaction}). The equivalence relation/coboundary transformation is also modified: we say that $\omega \sim \omega'$ if 
\begin{align}
\frac{\omega'(g,h,k)}{\omega(g,h,k)} = \frac{\nu(gh,k)\nu(g,h)}{\nu(g,hk)[g\nu(h,k)]}
\label{coboundaryanti}
\end{align}
where $\nu(g,h)\in U(1)$ and, again, $g\nu$ denotes the group action (\ref{gaction}).

\subsection{Outline of procedure}

The procedure for computing anomalies in the general case is exactly the same as the unitary case. As before, the first step is to choose an (edge) Hamiltonian $H$ that breaks the $G$-symmetry spontaneously and completely. This Hamiltonian $H$ has $|G|$ ground states which we label by $|\Omega; g\>$. The next step is to define domain wall states $|g_x\>$, and corresponding domain wall movement and splitting operators, $M^g_{x'x}$ and $S(g,h)$. Again, we use exactly the same definitions as in Sec.~\ref{sec:micro}.

The last step is to compute the $F$-symbol for the domain walls, $F(g,h,k)$. Again, we use the same definition as before, namely $F(g,h,k) = \<2|1\>$ where states $|1\>$ and $|2\>$ are defined as in Eq.~\ref{fdef1}.

The only new element in the antiunitary case involves the structure of the $F$-symbol. In particular, in the general antiunitary case, one can show that $F$ obeys the \emph{modified} cocycle condition (\ref{cocycleanti}), and is well-defined up to the \emph{modified} coboundary transformation (\ref{coboundaryanti}). As a result, $F$ is naturally an element of the cohomology group $H^3(G, U_T(1))$. For a derivation of the modified cocycle condition, see Appendix~\ref{pentagon}; we discuss the modified coboundary transformation in the next subsection.

\subsection{Checking the microscopic definition}

 In this section, we show that different choices of domain wall states or movement and splitting operators give the same $F$ up to the coboundary transformation (\ref{coboundaryanti}). This result, together with the derivation of the cocycle condition discussed in Appendix~\ref{pentagon}, guarantees that our procedure produces a well-defined element of $H^3(G, U_T(1))$.
 
A key result which we will need in our derivation is the following identity, which generalizes Eq.~\ref{sym_inv}. Let $\mathcal{O}$ be any operator that is invariant under all the symmetries and is supported near a domain wall $g$ at point $x$. We claim that
\begin{align}
\label{au_sym_inv}
\<...,g_x,...|\mathcal{O}|...,g_x,...\> &= g_L\<g_x|\mathcal{O}|g_x\>
\end{align}
where $g_L$ is the product of all domain walls to the left of $x$, and where the expression on the right hand side is defined by the group action (\ref{gaction}). To derive this identity, note that
\begin{align}
\<...,g_x,...|\mathcal{O}|...,g_x,...\> &= \<g_x; g_L|\mathcal{O}|g_x; g_L\> \nonumber \\
&= \< U^{g_L} g_x| \mathcal{O}| U^{g_L} g_x\> \nonumber \\
&= g_L \< g_x| \mathcal{O}| g_x\> 
\end{align}
Here, the first equality follows from (\ref{domain_wall_inv}), while the second equality follows from the definition, $|g_x; g_L\> = U^{g_L} |g_x\>$. The third equality follows from the fact that $\mathcal{O}$ commutes with $U^{g_L}$ together with the defining property of unitary/anti-unitary operators, namely $\<U^gv|U^gw\> = g\<v|w\>$.

With Eq.~(\ref{au_sym_inv}) in hand, we are now ready to show that $F$ is well-defined up to the coboundary transformation (\ref{coboundaryanti}). To start, let us see how $F$ transforms if we change the movement and splitting operators:
\begin{align}
M^g_{x'x} \rightarrow M'^g_{x'x}, \quad \quad S(g,h) \rightarrow S'(g,h)
\end{align}
Similarly to the unitary case, the definition of movement and splitting operators implies that
\begin{align}
M'^g_{x'x}|g_x\>&=e^{i\theta_{x'x}^g}M^g_{x'x}|g_x\> \nonumber \\
S'(g,h)|gh_1\>&=e^{i\phi(g,h)}S(g,h)|gh_1\>
\end{align}
for some real valued $\theta^g_{x'x}$ and $\phi(g,h)$. Likewise, for the multi-domain wall states,
\begin{align}
M'^g_{x'x} |...,g_x,...\> &= \omega_1 \cdot M^g_{x'x} |...,g_{x},...\> \label{omega1gen}\\
S'(g,h)|...,gh_1,...\> &= \omega_2 \cdot S(g,h)|...,gh_1,...\>
\label{omega2gen}
\end{align}
for some $U(1)$ phases, $\omega_1, \omega_2$. 
To find the relation between $\omega_1, \omega_2$ and $e^{i\theta^g_{x'x}}, e^{i\phi(g,h)}$, we multiply both sides of Eq.~(\ref{omega1gen}) by $\<...,g_x,...| (M^g_{x'x})^\dagger$, and then we apply (\ref{au_sym_inv}) to derive
\begin{align}
\omega_1 &=\<...,g_x,...|(M^g_{x'x})^\dagger M'^g_{x'x}|...,g_x,...\> \nonumber \\
&=g_{L,x}\<g_x|(M^g_{x'x})^\dagger M'^g_{x'x}|g_x\> \nonumber \\
&=g_{L,x}e^{i\theta^g_{x'x}}
\label{omega1form}
\end{align}
where $g_{L,x}$ is the product of all domain walls to the left of $x$. By the same reasoning, 
\begin{align}
\omega_2=g_{L,1}e^{i\phi(g,h)}
\label{omega2form}
\end{align}
where $g_{L,1}$ is the product of all domain walls to the left of the point $1$. 

Substituting (\ref{omega1form}) and (\ref{omega2form}) into (\ref{omega1gen}-\ref{omega2gen}), we derive
\begin{align}
M'^g_{x'x} |...,g_x,...\> &= (g_{L,x}e^{i\theta^g_{x'x}}) M^g_{x'x} |...,g_{x},...\>  \nonumber \\
S'(g,h)|...,gh_1,...\> &= (g_{L,1}e^{i\phi(g,h)}) S(g,h)|...,gh_1,...\> \label{omega12gen2}
\end{align}

Next, plugging (\ref{omega12gen2}) into the definition of $F$ (\ref{fdef1}-\ref{fdef2}), we obtain
\begin{align}
F'(g,h,k) = F(g,h,k)\frac{e^{i\phi(g,h)}e^{i\phi(gh,k)} \left(ge^{i\theta^h_{12}}\right)}{\left(ge^{i\phi(h,k)}\right)e^{i\phi(g,hk)}\left(ge^{i\theta^{hk}_{12}}\right)}
\end{align}
Crucially, this transformation matches the modified coboundary transformation (\ref{coboundaryanti}) with
\begin{align}
\nu(g,h)=e^{i\phi(g,h)}\left(ge^{i\theta^h_{12}}\right)
\end{align}
This is exactly what we want: different choices of movement and splitting operators lead to the same $F$, up to a modified coboundary transformation.

To complete the argument we also need to check that different choices of edge Hamiltonians and domain wall states lead to the same $F$, up to a coboundary transformation. Both properties follow by exactly the same reasoning as in the unitary case so we will not repeat it here.


\subsection{Example: chiral boson edge theory for $\mathbb{Z}_2\times\mathbb{Z}_2^T$ SPT phase}
\label{sec:z2z2example}
We now illustrate our general antiunitary procedure in an example. The example we consider is a continuum edge theory for a $\mathbb{Z}_2\times\mathbb{Z}_2^T$ bosonic SPT phase, discussed in Ref.~\onlinecite{LuVishwanath}. Here $\mathbb{Z}_2$ denotes a unitary symmetry and $\mathbb{Z}_2^T$ denotes an antiunitary symmetry. 

\subsubsection{Edge theory}

Similarly to the example discussed in Sec.~\ref{sec:z2field}, the edge theory is a chiral boson theory with two fields $\theta,\phi$ obeying the commutation relations
\begin{align*}
[\theta(x),\partial_y\phi(y)] &=2\pi i \delta(x-y) 
\end{align*}
with all other commutators vanishing. As in Sec.~\ref{sec:z2field}, the Hilbert space $\mathcal{H}$ is the usual infinite dimensional representation of the above algebra and the local operators in the edge theory are given by arbitrary derivatives and products of the elementary operators $\{e^{\pm i\theta}, e^{\pm i \phi}\}$.

In order to define the symmetry transformations, we first introduce some notation: we denote the elements of $G$ by  $\{1, s, t, st\}$, where $s$ is the (unitary) generator of $\mathbb{Z}_2$ and $t$ is the (antiunitary) generator of $\mathbb{Z}_2^T$. Likewise, we denote the $\mathbb{Z}_2\times\mathbb{Z}_2^T$ symmetry transformations by $\{U^1, U^s, U^t, U^{st}\}$. It suffices to specify the action of the two generators $U^s, U^t$. These generators act on $\theta, \phi$ as follows:
\begin{align}
U^s \theta (U^s)^{-1} &= \theta  \nonumber \\
U^s \phi (U^s)^{-1} &= \phi - \pi
\label{z2z2tsymm1k}
\end{align}
and
\begin{align}
U^t \theta (U^t)^{-1}  &= \theta - \pi \nonumber \\
U^t \phi (U^t)^{-1} &= -\phi 
\label{z2z2tsymm2k}
\end{align}
(Here, $U^s$ is unitary while $U^t$ is antiunitary).


\subsubsection{Calculating the anomaly}
To calculate the anomaly we need to choose an edge Hamiltonian that breaks the symmetry spontaneously and completely. We will do this in a slightly roundabout way: we will first introduce some auxiliary degrees of freedom into our edge theory. We will then break the symmetry of this enlarged edge theory. This approach will simplify our calculation.

To begin, consider the following chiral boson theory described by two fields $\bar{\theta}, \bar{\phi}$ obeying commutation relations
\begin{align*}
[\bar{\theta}(x),\partial_y\bar{\phi}(y)] &=2\pi i \delta(x-y) \end{align*}
with all other commutators vanishing. We assume that the symmetry acts on $\bar{\theta}, \bar{\phi}$ in the following way:
\begin{align}
U^s \bar{\theta} (U^s)^{-1} &= \bar{\theta}  \nonumber \\
U^s \bar{\phi} (U^s)^{-1} &= \bar{\phi} - \pi \nonumber \\
U^t \bar{\theta} (U^t)^{-1}  &= \bar{\theta} \nonumber \\
U^t \bar{\phi} (U^t)^{-1} &= -\bar{\phi}
\label{z2z2tsymm3k}
\end{align}

A crucial aspect of the $\bar{\theta}, \bar{\phi}$ chiral boson theory is that it is \emph{not} anomalous -- that is, it can be realized by a one dimensional lattice boson system with on-site $\mathbb{Z}_2\times\mathbb{Z}_2^T$ symmetry. One way to see this is to note that the above chiral boson theory is the standard low energy description of the 1D XXZ spin chain model where the $\mathbb{Z}_2^T$ symmetry $U^t$ is complex conjugation in the $\sigma^z$ basis, and the $\mathbb{Z}_2$ symmetry $U^s$ is the $\mathbb{Z}_2$ subgroup of the $U(1)$ symmetry of the XXZ model. Another way to see that the $\bar{\theta}, \bar{\phi}$ theory is not anomalous is to note that it can be gapped without breaking any the symmetries, for example by the Hamiltonian $H = H_{aux} - \int dx \ V \cos(\bar{\theta})$, where $H_{aux}$ is defined below.

Now, since the $\bar{\theta}, \bar{\phi}$ theory is not anomalous, we can add it to our edge theory without changing anything. (Physically, this corresponds to attaching a strictly 1D wire onto the edge of the SPT phase of interest). After enlarging our edge theory in this way, we then choose an edge Hamiltonian that breaks the $\mathbb{Z}_2\times\mathbb{Z}_2^T$ symmetry spontaneously and completely and opens up a gap. A Hamiltonian that does the job is
\begin{align}
H = H_0 + H_{aux} -  \int dx \ V [\cos (2\theta) + \cos (2 \bar{\phi})]
\end{align}
where $H_0, H_{aux}$ are the usual free boson Hamiltonians,
\begin{align*}
H_0&=\int dx \frac{1}{4\pi} [v_\theta (\partial_x \theta)^2 + v_\phi (\partial_x \phi)^2] \\
H_{aux}&=\int dx \frac{1}{4\pi} [v_{\bar{\theta}} (\partial_x \bar{\theta})^2 + v_{\bar{\phi}} (\partial_x \bar{\phi})^2]
\end{align*}
for some velocities $v_\theta, v_\phi, v_{\bar{\theta}}, v_{\bar{\phi}} > 0$.

For $V$ large and positive, the two cosine terms lock $\theta$ and $\bar{\phi}$ to one of two values $0, \pi$, spontaneously breaking the $\mathbb{Z}_2 \times \mathbb{Z}_2^T$ symmetry (\ref{z2z2tsymm1k}-\ref{z2z2tsymm3k}) and opening up an energy gap. 

We can now see the advantage of introducing the auxiliary fields $\bar{\theta}, \bar{\phi}$: these fields allow us to break the symmetry \emph{completely} in a simple way. If we dropped these fields, along with the corresponding cosine term, $\cos(2 \bar{\phi}(x))$, then the first term $V \cos(2\theta(x))$ would only break the $\mathbb{Z}_2^T$ symmetry and would leave the $\mathbb{Z}_2$ symmetry intact. Of course, we could also break the symmetry completely without introducing $\bar{\theta}, \bar{\phi}$, e.g. with a Hamiltonian of the form  $H = H_0 + \int dx V \cos(4 \phi)$, but this leads to a more complicated calculation.

Turning back to the calculation, we now discuss the ground states of $H$. As in the example in Sec.~\ref{sec:z2field}, we will take the limit $V \rightarrow \infty$ for simplicity. In this limit, the four ground states of $H$ are eigenstates of $e^{i \theta(x)}$ and $e^{i \bar{\phi}(x)}$ with eigenvalues $\pm 1$. We define $|\Omega; 1\>$ to be the state with eigenvalues $+1$:
\begin{align}
e^{i\theta(x)}|\Omega; 1\> &= |\Omega; 1\> \nonumber \\
e^{i\bar{\phi}(x)}|\Omega; 1\> &= |\Omega; 1\> 
\end{align}
We then define the other ground states by $|\Omega; g\> = U^g |\Omega; 1\>$, which corresponds to the following eigenvalue assignments (given Eqs.~\ref{z2z2tsymm1k}-\ref{z2z2tsymm3k}):
\begin{align}
e^{i\theta(x)}|\Omega; g\> &= e^{i2 \pi \lambda(g)} |\Omega; g\> \nonumber \\
e^{i\bar{\phi}(x)}|\Omega; g\> &= e^{i2 \pi \mu(g)} |\Omega; g\> 
\end{align}
where
\begin{equation}
\lambda(g)=\begin{cases} 
      0 & \quad g=1,s\\
      1/2 & \quad g=t, st
   \end{cases}
\end{equation}
and
\begin{equation}
\mu(g)=\begin{cases} 
      0 & \quad g=1,t\\
      1/2 & \quad g=s, st
   \end{cases}
\end{equation}

Next we define domain wall states $|g_x\>$ by:
\begin{align}
|g_x\>=(a^g_x)^\dagger|\Omega; 1\>
\end{align}
where the operators $(a^g_x)^\dagger$ are defined as
\begin{align}
(a^g_x)^\dagger &=e^{-i \int_x^\infty dy  \ (\lambda(g) \partial_y \phi(y) + \mu(g) \partial_y \bar{\theta}(y))} 
\end{align}

To see that $|g_x\>$ is a valid domain wall state, note that
\begin{align}
e^{i \theta(x')} |g_x\> = \begin{cases}  \ |g_x\> \ & x' < x \nonumber \\
e^{i 2\pi \lambda(g)} \ |g_x\> \ & x' > x
\end{cases}
\end{align}
and
\begin{align}
e^{i \bar{\phi}(x')} |g_x\> = \begin{cases}  \ |g_x\> \ & x' < x \nonumber \\
e^{i 2\pi \mu(g)} \ |g_x\> \ & x' > x
\end{cases}
\end{align}
by the commutation relations between $\theta, \phi$ and $\bar{\theta}, \bar{\phi}$. 

The next step is to define movement and splitting operators. We define the movement operator $M^g_{x' x}$, for $x' > x$, by
\begin{align}
M^g_{x' x}= \sum_{k \in G} U^k X^g_{x'x} P_{x'}^g (U^k)^{-1}, \quad \quad x' > x
\label{mdefz2z2}
\end{align}
where 
\begin{align}
X^g_{x'x}=e^{i \int_x^{x'} dy  \ (\lambda(g) \partial_y \phi(y) + \mu(g) \partial_y \bar{\theta}(y))}
\end{align}
and 
\begin{align}
    P_{x'}^g = \frac{(1+e^{i \theta(x')} e^{-i 2\pi \lambda(g)})(1+e^{i \bar{\phi}(x')} e^{-i 2\pi \mu(g)})}{4}
\end{align}
Here $P_{x'}^g$ should be thought of as a local ground state projection operator, supported near $x'$. The defining property of this operator is that it leaves invariant the state $|\Omega; g\>$ and annihilates the other ground states: that is, $P_{x'}^g |\Omega; h\> = \delta_{gh} |\Omega; g\>$.

To see that $M^g_{x' x}$ is a valid movement operator, note that it is local and $\mathbb{Z}_2\times\mathbb{Z}_2^T$ symmetric by construction. We can also see that it has the correct action on domain wall states: 
\begin{align}
    M^g_{x'x} |g_x\> = X^g_{x'x} |g_x\> = |g_{x'}\>
\end{align}
Here, the first equality follows from the observation that only the $k=1$ term in (\ref{mdefz2z2}) gives a nonzero result when acting on $|g_x\>$, while the second equality follows from the fact that $X^g_{x'x} (a_x^g)^\dagger = (a_{x'}^g)^\dagger$.

Likewise, we define the reverse movement operator by 
\begin{align}
M^g_{x x'}= (M^g_{x'x})^\dagger, \quad \quad x' > x
\end{align}

Moving on to splitting operators, we define $S(g,h)$ by
\begin{align}
S(g,h) &= \sum_{k \in G} U^k U^g X^h_{21} (U^g)^{-1} C(g,h) P_{2}^{gh} (U^k)^{-1}
\label{sdefz2z2}
\end{align}
where
\begin{align}
C(g,h) = e^{i [p(g,h) \phi(1) + q(g,h)  \bar{\theta}(1)]} 
\label{cghdefk}
\end{align}
and
\begin{align}
p(g,h) &=\lambda(g)+\lambda(h)-\lambda(gh) \nonumber \\
q(g,h) &=\mu(g)+\sigma(g) \mu(h)-\mu(gh)
\end{align}
Here $\sigma(g) = \pm 1$ depending on whether $g$ is unitary or antiunitary, i.e. $\sigma(g) = 1 - 4\lambda (g)$. 

To see that $S(g, h)$ is a valid splitting operator, notice that it is local and $\mathbb{Z}_2\times\mathbb{Z}_2^T$ symmetric by construction. We can also check that $S(g, h)$ has the correct action on domain walls:
\begin{align}
S(g, h)|gh_1\> &=  U^g X^h_{21} (U^g)^{-1} C(g,h)|gh_1\> \nonumber \\
&\propto U^g (a^h_2)^\dagger (U^g)^{-1} (a^g_1)^\dagger |\Omega;1\> \nonumber \\
&\propto |g_1, h_2\>
\end{align}
Here, the first equality follows from the fact that only the $k=1$ term in (\ref{sdefz2z2}) gives a nonzero result when acting on $|gh_1\>$, while the second equality follows from the definition of $(a^g_x)^\dagger$. The third equality follows by noting that $U^g (a^h_2)^\dagger (U^g)^{-1} (a^g_1)^\dagger |\Omega;1\>$ has the same local expectation values as $|g_1\>$ near $x=1$, and the same local expectation values as $U^g|h_2\>$ near $x=2$, and the same local expectation values as $U^g |\Omega;1\>$ between $x=1$ and $x=2$. 

With these operators defined, we are ready to compute $F(g, h, k)$.  From equation \ref{fdef1},
\begin{align}
|1\> &=  [U^g X^h_{12} (U^g)^{-1}] \cdot X^g_{01} \cdot [U^g X^h_{21} (U^{g})^{-1}] \nonumber \\
&\cdot C(g,h) \cdot [U^{gh} X^{k}_{32}X^{k}_{21} (U^{gh})^{-1}] \cdot C(gh,k)|ghk_1\> \nonumber \\
|2\> &= [U^{gh} X^k_{32}X^k_{21} (U^{gh})^{-1}] \cdot [U^g C(h,k) (U^g)^{-1}] 
\nonumber \\
&\cdot [U^g X^{hk}_{12} (U^g)^{-1}] \cdot X^g_{01} \cdot [U^g X^{hk}_{21} (U^g)^{-1}] \nonumber \\
&\cdot C(g, hk)|ghk_1\>
\end{align}

In order to compare these two states, we note that all the operators in the above products commute with one another. Using this commutativity together with the identity $X^\alpha_{xx'}=(X^\alpha_{x'x})^{-1}$, we can rewrite our states as:
\begin{align}
|1\> &=X^g_{01} U^{gh} X^{k}_{32} X^{k}_{21} (U^{gh})^{-1} \nonumber \\
&\cdot C(g,h)  C(gh,k)|ghk_1\> \nonumber \\
|2\> &= X^g_{01} U^{gh} X^{k}_{32} X^{k}_{21} (U^{gh})^{-1} \nonumber \\
&\cdot U^g C(h,k) (U^g)^{-1} C(g, hk)|ghk_1\>
\end{align}

Next, using the definition (\ref{cghdefk}) and the symmetry action, one can check that
\begin{align}
&C(g,h)  C(gh,k) |ghk_1\>  = \nonumber \\
& \quad e^{i 2\pi\mu(g) p(h,k)} U^g C(h,k) (U^g)^{-1} C(g,hk)|ghk_1\>  
\end{align}

We conclude that $|1\> = e^{i 2\pi \mu(g) p(h,k)}|2\>$ so that
\begin{align}
F(g,h,k) &= \<2|1\> =  e^{i 2\pi\mu(g) p(h,k)}
\end{align}

Having computed $F$, the next question is to determine whether $F$ is a trivial or nontrivial cocycle, i.e. a trivial or non-trivial element of $H^3(G,U_T(1))$. To answer this question, we compute the following two gauge invariant combinations of $F$: 
\begin{align}
F(s,s,s)F(s,1,s) &=1 \nonumber \\
\chi_s(t,t)\chi_s(1,t) &=-1
\end{align}
where
\begin{align}
\chi_g(h,k)=\frac{F(g,h,k)F(h,k,g)}{F(h,g,k)}
\end{align}
Given that the second quantity is different from $1$, it follows that $F$ is a nontrivial cocycle. (Both quantities are $1$ for a trivial cocycle). We conclude that our edge theory describes the boundary of one of the three nontrivial SPT phases within the $\mathcal{H}(\mathbb{Z}_2\times \mathbb{Z}_2^T,U(1)_T)=\mathbb{Z}_2\times \mathbb{Z}_2$ classification~\cite{SPTCo}, which is consistent with the original analysis of Ref.~\onlinecite{LuVishwanath}.

\section{Continuous symmetries}
\label{sec:continuous}
\subsection{Outline of procedure}

So far we have presented a procedure for computing anomalies for SPT edge theories with a discrete symmetry group $G$. We now discuss how to generalize this procedure to edge theories with \emph{continuous} symmetries.

The main obstruction to applying our procedure in the continuous case has to do with the definition of domain wall states. Recall that in the discrete case we construct domain walls as follows: first, we choose an (edge) Hamiltonian that breaks the symmetry spontaneously and completely and opens up an energy gap at the edge. Such a Hamiltonian has a collection of degenerate ground states, which we label by $|\Omega; g\>$. We then define domain wall excitations to be states that have the same local expectation values as one ground state, $|\Omega; g\>$, in some large interval, and the same local expectation values as another ground state $|\Omega; h\>$ in a neighboring large interval. The problem with applying this scheme to continuous symmetries is that it is impossible to both break a continuous symmetry \emph{and} open up an energy gap: spontaneously breaking a continuous symmetry always leads to gapless Goldstone modes.

The key to overcoming this problem is to recognize that we don't actually need a Hamiltonian with spontaneous symmetry breaking and an energy gap: this symmetry-breaking Hamiltonian provides a nice physical context for thinking about domain wall states -- our main objects of interest -- but it isn't strictly necessary for our procedure. From an operational point of view, all that we need is a single (edge) state $|\Psi\>$ with two properties: (i) $|\Psi\>$ breaks the symmetry explicitly and completely -- i.e. is not invariant under any of the symmetry transformations; and (ii) $|\Psi\>$ is the unique ground state of a gapped, local Hamiltonian. (The latter condition is important to guarantee that $|\Psi\>$ has various locality properties like short-range correlations). Once we have such a state, we denote it by $|\Psi\> \equiv |\Omega; 1\>$ and then define $|\Omega; g\>$ by
\begin{align}
|\Omega;g\>=U^g|\Omega;1\>
\end{align}
Having defined the $|\Omega; g\>$ states, we then define our domain wall states and calculate $F(g,h,k)$ exactly as before. This modified procedure works equally well for either continuous or discrete symmetry groups.

One subtlety that appears in the continuous symmetry case is that SPT phases with continuous symmetry groups are conjectured to be classified by the \emph{Borel} cohomology group $\mathcal{H}^3_\mathcal{B}(G,U_T(1))$~\cite{SPTCo}. This means that if we want to interpret $F(g,h,k)$ as a label for a bulk SPT phase, we need to show that $F(g,h,k) \in\mathcal{H}^3_\mathcal{B}(G,U_T(1))$. This amounts to showing that (i) $F(g,h,k)$ is a Borel measurable 3-cocycle, and (ii) $F(g,h,k)$ is well-defined up to multiplication by the coboundary of a Borel measurable 2-cocycle. Neither of these properties are obvious from our definition of $F$. Indeed, if we do not impose any additional constraints on the splitting operators $S(g,h)$ and the movement operators $M^g_{x'x}$, then we can only show weaker versions of (i) and (ii) that do not include the Borel measurability constraints. In this paper, we will not attempt to fill in this gap, but we expect that both properties (i) and (ii) follow naturally if the splitting operators $S(g,h)$ and movement operators $M^g_{x'x}$ are chosen so that their dependence on $g,h \in G$ is not too discontinuous; for example, it may be enough to require that the matrix elements of $S(g,h)$ and $M^g_{x'x}$ are piecewise continuous as a function of $g$ and $h$. We will see evidence for this conjecture in the next example: there we will see that a reasonable choice of splitting and movement operators leads to a Borel measurable $F(g,h,k)$.

\subsection{Example: chiral boson edge theory for bosonic IQH phase}
\label{sec:IQH_example}
We now demonstrate our approach with an example: an edge theory for the bosonic integer quantum Hall state -- a nontrivial bosonic SPT phase with $U(1)$ symmetry. This edge theory was introduced in Ref.~\onlinecite{LuVishwanath}.

\subsubsection{Edge theory}
Like the examples discussed in Sec.~\ref{sec:z2field} and Sec.~\ref{sec:z2z2example}, the edge theory we consider is a chiral boson edge theory with two fields $\theta,\phi$ obeying the commutation relations
\begin{align*}
[\theta(x),\partial_y\phi(y)] &=2\pi i \delta(x-y) 
\end{align*}
with all other commutators vanishing. As in the previous examples, the local operators in the edge theory are given by arbitrary products and derivatives of the elementary operators $\{e^{\pm i \theta}, e^{\pm i \phi}\}$ and the Hilbert space $\mathcal{H}$ is the usual infinite dimensional representation of the above algebra. 

To complete the edge theory, we need to specify the $U(1)$ symmetry transformation. Denoting these transformations by $U^\alpha$, where $\alpha \in [0, 2\pi)$, the symmetry action is given by:
\begin{align}
      U^\alpha \theta (U^\alpha)^{-1} &= \theta - \alpha \nonumber \\
      U^\alpha \phi (U^\alpha)^{-1} &= \phi - \alpha
\label{U1symm}
\end{align}
We note that this edge theory reduces to the one discussed in Sec.~\ref{sec:z2field} if we restrict to the $\mathbb{Z}_2$ subgroup of $U(1)$.

\subsubsection{Calculating the anomaly}

The first step in calculating the anomaly is to choose a state $|\Psi\> \equiv |\Omega;0\>$\footnote{We denote the identity element by $0$ since we are using additive notation.} that (i) breaks the $U(1)$ symmetry completely and (ii) is the unique ground state of a gapped local Hamiltonian. To this end, we define $|\Omega; 0\>$ to be the (unique) simultaneous eigenstate of the operators $e^{i\theta(x)}$ with eigenvalue $1$:
\beq
e^{i\theta(x)}|\Omega;0\>=|\Omega;0\>
\eeq
for all $x$. To see that $|\Omega ; 0\>$ has the required properties, note that it breaks the $U(1)$ symmetry defined in (\ref{U1symm}), and furthermore it is the unique ground state of the (gapped) Hamiltonian $H_0 - \int dx \ V \cos(\theta)$ in the limit $V \rightarrow \infty$, where $H_0 = \int dx \frac{1}{4\pi} [v_\theta (\partial_x \theta)^2 + v_\phi (\partial_x \phi)^2]$.

After choosing $|\Omega; 0\>$, we then define the remaining vacuum states $|\Omega;\alpha\>$ to be symmetry partners of $|\Omega; 0\>$.
\begin{align}
|\Omega; \alpha\> = U^\alpha |\Omega; 0\>
\end{align}
By construction $|\Omega; \alpha\>$ is a simultaneous eigenstate of $e^{i\theta(x)}$ with eigenvalue $e^{i\alpha}$:
\beq
e^{i\theta(x)}|\Omega;\alpha\>= e^{i \alpha} |\Omega;\alpha\>
\eeq

The next step is to define domain wall states $|\alpha_x\>$ that spatially interpolate between the two states, $|\Omega ; 0\>$
and $|\Omega; \alpha\>$. We define 
\begin{align}
|\alpha_x\> = (a^\alpha_x)^\dagger|\Omega;1\>
\end{align}
where $(a^\alpha_x)^\dagger$ is defined by
\begin{align}
(a^\alpha_x)^\dagger=e^{-i\frac{\alpha}{2\pi}\int_x^\infty dy \ \partial_y\phi(y)}
\end{align}
We can see that $|\alpha_x\>$ is a valid domain wall state since
\begin{align}
e^{i \theta(x')} |\alpha_x\> = \begin{cases}  \ |\alpha_x\> \ & x' < x \nonumber \\
e^{i \alpha} \ |\alpha_x\> \ & x' > x
\end{cases}
\end{align}
by the commutation relations between $\theta$ and $\phi$. 

Next, we define movement and splitting operators for the above domain walls. We define the movement operator by
\begin{align}
M^{\alpha}_{x'x}=e^{i\frac{\alpha}{2\pi}\int_x^{x'} dy \ \partial_y\phi(y)}
\end{align}
To see that this is a valid movement operator, note that $M^{\alpha}_{x'x}$ is local and $U(1)$ symmetric by construction. Also,  $M^{\alpha}_{x'x}$ has the correct action on domain walls: $M^{\alpha}_{x'x}|\alpha_x\>\propto|\alpha_{x'}\>$ since $M^{\alpha}_{x'x}(a^\alpha_x)^\dagger\propto (a^\alpha_{x'})^\dagger$.  

Moving on to splitting operators, we define $S(\alpha,\beta)$ by
\begin{align}
S(\alpha,\beta)&=M^\beta_{21}C(\alpha,\beta) \nonumber \\
&=e^{i\frac{\beta}{2\pi}\int_1^{2} dy \ \partial_y\phi(y)}C(\alpha,\beta)
\end{align}
where $C(\alpha,\beta)$ denotes the operator
\beq
C(\alpha,\beta)=e^{\frac{i}{2\pi}(\alpha + \beta - [\alpha + \beta])(\phi(1)-\theta(1^-))}
\label{Cabdef}
\eeq
Here we use the symbol $[x]$ to denote the unique number in $[0,2 \pi)$ that is equal to $x$ modulo $2\pi$. Also, the notation $\theta(1^-)$ is shorthand for $\theta(1-\epsilon)$ where $\epsilon$ is a small positive number. (It is important to carefully distinguish between $\theta(1^-)$ and $\theta(1^+)$ because we will be applying this operator to a domain wall state with a $\theta$ domain wall at $x=1$.)

To see that $S(\alpha, \beta)$ is a valid splitting operator, notice that it is local and $U(1)$ symmetric by construction. We can also see that $S(\alpha, \beta)$ has the correct action on domain walls:
\begin{align}
S(\alpha,\beta)|[\alpha+\beta]_1\> &= M^\beta_{21} C(\alpha,\beta) |[\alpha+\beta]_1\> \nonumber \\
&= M^\beta_{21} e^{\frac{i}{2\pi}(\alpha + \beta - [\alpha + \beta])\phi(1)}|[\alpha+\beta]_1\> \nonumber \\
&\propto (a^\beta_2)^\dagger (a^\alpha_1)^\dagger |\Omega;1\> \nonumber \\
&\propto |\alpha_1, \beta_2\>
\end{align}
Here, the second equality follows from the fact that $e^{i \theta(1^-)} |[\alpha+\beta]_1\> = |[\alpha+\beta]_1\>$, while the third equality follows from the definition of $(a_x^\alpha)^\dagger$. The last equality follows from the fact that the state $(a^\beta_2)^\dagger (a^\alpha_1)^\dagger |\Omega;1\>$ has the same expectation values for local operators as $|\alpha_1, \beta_2\>$.

With these operators defined, we are ready to compute $F(\alpha,\beta,\gamma)$.  From equation \ref{fdef1},
\begin{align}
|1\>=&M^\beta_{12}M^\alpha_{01}M^\beta_{21}C(\alpha,\beta)M^{\gamma}_{32}M^{\gamma}_{21} \nonumber \\
&\cdot C([\alpha+\beta],\gamma)|[\alpha+\beta+\gamma]_1\> \nonumber \\
|2\>=&M^\gamma_{32}M^\gamma_{21}C(\beta,\gamma)M^{[\beta+\gamma]}_{12}M^\alpha_{01}M^{[\beta+\gamma]}_{21} \nonumber \\
&\cdot C(\alpha,[\beta+\gamma])|[\alpha+\beta+\gamma]_1\>
\end{align}

In order to compare these two states, we will reorder the operators with the following identities derived using the Baker-Campbell-Hausdorff formula: 
\begin{align}
M^\alpha_{x'x} M^\beta_{y'y}  &= M^\beta_{y'y} M^\alpha_{x'x} \\
C(\beta, \gamma) M^\alpha_{x1} &= e^{-\frac{i\alpha}{2\pi} (\beta + \gamma - [\beta + \gamma]) \Theta(1-x)} M^\alpha_{x1} C(\beta, \gamma)
\nonumber 
\end{align}
where $\Theta(x)$ denotes the Heaviside step function. With these formulas and the identity $M^\alpha_{xx'}=(M^\alpha_{x'x})^{-1}$, we can rewrite our states as:
\begin{align}
|1\> &=M^\alpha_{01}M^{\gamma}_{32}M^{\gamma}_{21} C(\alpha,\beta)C([\alpha+\beta],\gamma)|[\alpha+\beta+\gamma]_1\>\nonumber\\
|2\> &= e^{i \Gamma}M^\alpha_{01}M^\gamma_{32}M^\gamma_{21} C(\beta,\gamma)C(\alpha,[\beta+\gamma])|[\alpha+\beta+\gamma]_1\>
\end{align}
where $\Gamma =  - \frac{\alpha}{2\pi}(\beta + \gamma - [\beta + \gamma])$. 

Next, using the definition (\ref{Cabdef}), it is easy to check that
\begin{align}
C(\alpha, \beta)  C([\alpha + \beta], \gamma)  = C(\beta, \gamma) C(\alpha,[\beta + \gamma])  
\end{align}
Substituting this identity into the above expression for $|1\>, |2\>$, we see that $|2\> = e^{i \Gamma} |1\>$.
We conclude that
\begin{align}
F(\alpha,\beta,\gamma)= \<2|1\> =e^{ \frac{i\alpha}{2\pi} (\beta + \gamma - [\beta + \gamma])}
\end{align}
Note that $F$ is a piecewise continuous function of $\alpha, \beta, \gamma$ and is therefore Borel measurable. This is consistent with our conjecture that $F$ will always be Borel measurable for piecewise continuous movement and splitting operators. 

Having computed $F$, the next question is to determine whether $F$ is a trivial or non-trivial cocycle. To answer this question, we compute the following gauge invariant quantity:
\begin{align}
F(\pi,\pi,\pi) F(\pi,0,\pi) = -1
\end{align}
Since this quantity is different from $1$, it follows that $F$ is a non-trivial cocycle. We conclude that our edge theory describes the boundary of a nontrivial bosonic SPT phase with $U(1)$ symmetry. This is consistent with previous work~\cite{LuVishwanath}.



\section{Conclusion}
\label{sec:conclusion}

In this paper, we have presented a general procedure for calculating anomalies in (1D) bosonic SPT edge theories. Our procedure takes as input a bosonic SPT edge theory and produces as output an element $\omega \in H^3(G,U_T(1))$ describing the anomaly carried by the edge theory. An important feature of our procedure is that, unlike previous approaches, it applies to general bosonic SPT edge theories with both unitary and antiunitary symmetries, with the only restriction being that the underlying 2D symmetry must be on-site. 

One class of SPT edge theories that we cannot analyze with our current approach are those with spatial symmetries such as translation or reflection symmetries~\cite{point_group, space}. The problem is that, for these types of edge theories, it is impossible to construct movement and splitting operators with the two requirements that they are are local in space and also invariant under all the symmetries. A potential hint for how to overcome this problem, at least in some cases, is the observation~\cite{point_group} that SPT phases with point group symmetries are closely connected to SPT phases with on-site symmetries in fewer spatial dimensions. This suggests that our method might be suitable for studying boundary theories of higher dimensional (e.g. 3D) SPT phases with point group symmetries.

Our anomaly calculation always starts by choosing an edge Hamiltonian that opens up a gap and breaks the symmetry on the edge completely. However, in some edge theories, it is possible to open up a gap by only breaking \emph{some} of the symmetries. (A famous example is the 2D topological insulator: in this case the edge can be gapped by breaking time-reversal symmetry while preserving $U(1)$ charge conservation symmetry). It would be interesting to develop methods for computing anomalies in this partial symmetry-breaking scenario. In this case, domain walls have more structure: they can carry quantum numbers under the symmetry, in addition to their fusion properties. We expect that the anomaly is encoded in this more complicated set of data. This problem may be related to the decorated domain wall construction of Ref.~\onlinecite{partial}.

Another interesting direction for future work would be to generalize our approach to (1D) \emph{fermionic} SPT edge theories. The fermionic case is especially intriguing given that there are 2D fermionic SPT phases that are beyond~\cite{ChengGufSPT} the supercohomology classification scheme~\cite{gu2014symmetry}. Despite the complexity of 2D fermionic SPT phases, we expect that our basic approach is still applicable: that is, given any fermionic SPT edge theory, we can determine the identity of the corresponding 2D SPT phase by breaking the symmetry and then extracting the fusion rules and $F$-symbol of the domain walls at the edge.


\acknowledgments

We thank Colin Aitken for useful discussions. K.K. and M.L. acknowledge the support of the Kadanoff Center for Theoretical Physics at the University of Chicago.  This work was supported in part by the Simons Collaboration on Ultra-Quantum Matter, which is a grant from the Simons Foundation (651440, ML).


\begin{appendix}


\section{Cocycle condition}
\label{pentagon}
 
\begin{figure}[tb]
\centering
\includegraphics[width=1.0\columnwidth]{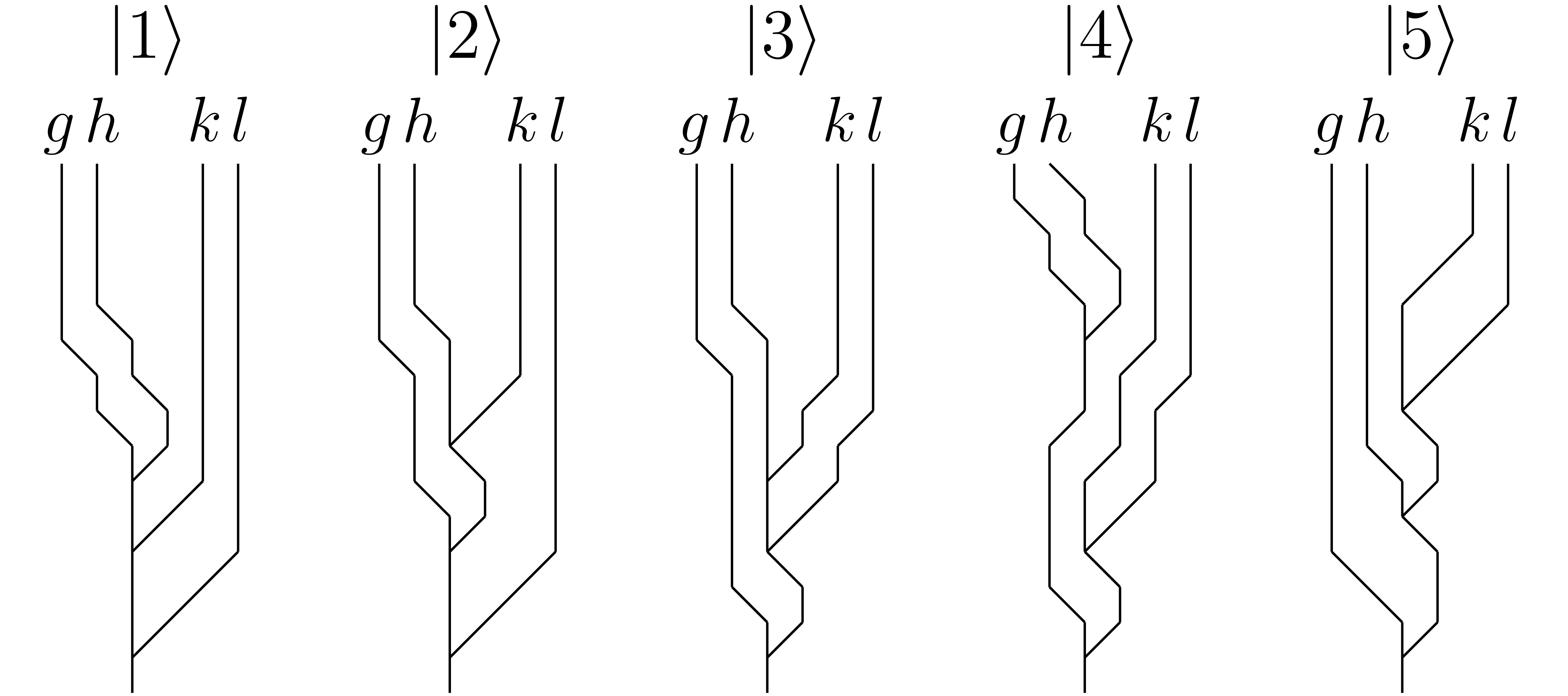}
\caption{The five microscopic states used to prove the cocycle condition.}
\label{fig:pent_states}
\end{figure}

In this appendix, we show that the domain wall $F$-symbol defined in (\ref{fdef1}-\ref{fdef2}) is a cocycle. More specifically, we show that $F$ obeys Eq.~\ref{cocycleanti} -- the general cocycle condition for symmetry groups containing both unitary and antiunitary symmetries.

Our proof closely follows the derivation of the pentagon identity for anyonic $F$-symbols, presented in Appendix A of Ref.~\onlinecite{KL}. 
As in Ref.~\onlinecite{KL}, the first step is to pick a nice phase convention for the movement operators $M^g_{x'x}$. Specifically, we choose the phases of the movement operators so that
\begin{align}
M^g_{xx'}M^g_{x'x}|g_x\> &= |g_x\> \nonumber \\
M^g_{x''x'}M^g_{x'x}|g_x\> &= M^g_{x''x}|g_{x}\>
\label{mid}
\end{align}
With this phase convention, it is possible to show that our space-time diagrams satisfy the following topological invariance property: consider any process, $P$, composed out of a sequence of movement and splitting operators acting on an initial state 
\begin{align} 
|i\> = |..., g_x, h_{x'}, k_{x''},...\>. 
\end{align}
For any such process, we can draw a corresponding space-time diagram. Next, consider a second process, $P'$, that acts on the same initial state, $|i\>$, and that leads to the same final state. Again, we can draw a corresponding space-time diagram. The topological invariance property says that if these two space-time diagrams can be continuously deformed into one another while fixing the endpoints, then the two processes produce the same final states with the same phases. That is, 
\begin{align}
P |i\> = P' |i\>
\label{topinvprop}
\end{align}
The proof of the topological invariance property (\ref{topinvprop}) is identical to the one given in Appendix A of Ref.~\onlinecite{KL}.

With the help of the topological invariance property (\ref{topinvprop}), we will now show that the $F$-symbol defined in (\ref{fdef1}-\ref{fdef2}) satisfies the cocycle condition (\ref{cocycleanti}). Consider the five processes shown in Fig.~\ref{fig:pent_states}. Let us denote the final states of these processes by $|1\>, |2\>, |3\>, |4\>, |5\>$. Notice that these states are the same up to a phase since they describe the same four domain walls in the same four positions. The idea of the proof is to compute the phase difference between state $|1\>$ and state $|5\>$ in two difference ways. More specifically, using the topological invariance property (\ref{topinvprop}) we will show that
\begin{align}
|1\>&=F(g,h,k)|2\>\nonumber\\
|2\>&=F(g,hk,l)|3\>\nonumber\\
|3\>&=g (F(h,k,l))|5\>
\end{align}
and
\begin{align}
|1\>&=F(gh,k,l)|4\>\nonumber\\
|4\>&=F(g,h,kl)|5\>
\end{align}
where the $g$ action is defined as in Eq.~\ref{gaction}.
Putting this all together gives us the desired cocycle condition:
\begin{align}
\frac{F(g,h,k)F(g,hk,l)[gF(h,k,l)]}{F(gh,k,l)F(g,h,kl)} = 1
\end{align}

We now derive each of the above equations.
To aid in the discussion, we define the operators
\begin{align}
\mathcal{O}_1(g,h,k)&=M^h_{12}M^g_{01}S(g,h)M^k_{32}S(gh,k)\nonumber\\
\mathcal{O}_2(g,h,k)&=M^k_{32}S(h,k)M^{hk}_{12}M^g_{01}S(g,hk)
\end{align}

To derive the first equation, $|1\>=F(g,h,k)|2\>$, notice that processes $1$ and $2$ start in the same state $|ghkl_1\>$, and they also contain the same sequence of movement and splitting operators starting from the beginning of the process, leading up to the (intermediate) state $|ghk_1,l_4\>$. It is at that point that the two processes diverge: in process $1$, the operator $\mathcal{O}_1(g,h,k)$ is applied while in process $2$, the operator $\mathcal{O}_2(g,h,k)$ is applied. After that, the two processes again coincide. It follows that the phase difference between $|1\>$ and $|2\>$ comes entirely from the difference between $\mathcal{O}_1$ and $\mathcal{O}_2$. That is,
\begin{align}
\<2|1\>&=\<ghk_1,l_4|\mathcal{O}_2(g,h,k)^\dagger\mathcal{O}_1(g,h,k)|ghk_1,l_4\>\nonumber\\
&=\<ghk_1|\mathcal{O}_2(g,h,k)^\dagger\mathcal{O}_1(g,h,k)|ghk_1\>\nonumber\\
&=F(g,h,k)
\end{align}
where the second equality follows from (\ref{au_sym_inv}) -- one of the fundamental properties of multi-domain wall states.

The second equation, $|2\>=F(g,hk,l)|3\>$, follows from similar logic. In this case, we use the topological invariance property (\ref{topinvprop}) to redraw process $2$ so that it is identical to process $3$ except for an $F$-move at the very beginning of the process (see Fig. 20 of Ref.~\onlinecite{KL}). After this modification, the only difference between the two processes is that in process $2$ the operator $\mathcal{O}_1(g,hk,l)$ is applied at the beginning, while in process $3$, the operator $\mathcal{O}_2(g,hk,l)$ is applied. It then follows that
\begin{align}
\<3|2\>&=\<ghkl_1|\mathcal{O}_2(g,hk,l)^\dagger\mathcal{O}_1(g,hk,l)|ghkl_1\>\nonumber\\
&=F(g,hk,l)
\end{align}
The third equation, $|3\>=g(F(h,k,l))|5\>$ is the trickiest one, and the one that distinguishes the unitary and antiunitary cases. Again, we use the topological invariance property to redraw process $3$ so that it differs from process $5$ by a replacement $\mathcal{O}_1(h,k,l) \rightarrow \mathcal{O}_2(h,k,l)$ (see Fig. 21 of Ref.~\onlinecite{KL}). Hence 
\begin{align}
\<5|3\>&=\<g_{-1},hkl_1|\mathcal{O}_2(h,k,l)^\dagger\mathcal{O}_1(h,k,l)|g_{-1},hkl_1\>\nonumber\\
&=g\<hkl_1|\mathcal{O}_2(h,k,l)^\dagger\mathcal{O}_1(h,k,l)|hkl_1\>\nonumber\\
&=gF(g,h,k)
\end{align}
where the second equation follows from (\ref{au_sym_inv}). Note that the key difference between this equation and the others is that the $F$-move takes place to the right of the $g$ domain wall; this is the origin of the $g$ action.

The other two equations, $|1\>=F(gh,k,l)|4\>$ and $|4\>=F(g,h,kl)|5\>$, follow from similar reasoning. In both cases, the relevant processes can be related by topological invariance together with an appropriate $F$-move (see Figs. 22 and 23 of Ref.~\onlinecite{KL}). We then derive the equations as before:
\begin{align}
\<4|1\>&=\<ghkl_1|\mathcal{O}_2(gh,k,l)^\dagger\mathcal{O}_1(gh,k,l)|ghkl_1\>\nonumber\\
&=F(gh,k,l)\nonumber\\
\<5|4\>&=\<ghkl_1|\mathcal{O}_2(g,h,kl)^\dagger\mathcal{O}_1(g,h,kl)|ghkl_1\>\nonumber\\
&=F(g,h,kl)
\end{align}
This completes our derivation of the cocycle condition (\ref{cocycleanti}).



\end{appendix}


\bibliography{bosonic_SPT}
%

\end{document}